\documentclass{emulateapj}
\usepackage{amsmath,amssymb}
\usepackage{graphicx}

\newcommand{\ha}{H$\alpha$}
\newcommand{\has}{H$\alpha~$}
\newcommand{\hb}{H$\beta~$}
\newcommand{\oiii}{[O {\footnotesize III}]($\lambda$5007)}
\newcommand{\nii}{[N {\footnotesize II}]($\lambda$6583)}
\newcommand{\sii}{[S {\footnotesize II}]($\lambda\lambda$6716,6731)}

\newcommand{\msp}{$~$}

\newcommand{\iion}[2]{\hbox{#1\,{\sc #2}}}

\slugcomment{Accepted for publication in the Astrophysical Journal}

\shorttitle{Shock--ionized Gas in Nearby Starburst Galaxies}
\shortauthors{Hong et al.}

\begin{document}

\title{Constraining Stellar Feedback: Shock--ionized Gas in Nearby Starburst Galaxies}

\author{ 
Sungryong Hong\altaffilmark{1,2},
Daniela Calzetti\altaffilmark{1},
John S. Gallagher III\altaffilmark{3},
Crystal L. Martin\altaffilmark{4},
Christopher J. Conselice\altaffilmark{5}, 
and   
Anne Pellerin\altaffilmark{6}
}

\altaffiltext{1}{Department of Astronomy, University of Massachusetts,  
Amherst, MA 01003, USA}
\altaffiltext{2}{National Optical Astronomy Observatory, 
950 North Cherry Avenue, Tucson, AZ 85719, USA}
\altaffiltext{3}{Department of Astronomy, University of WisconsinÐMadison, 
475 North Charter Street, Madison, WI 53706, USA}
\altaffiltext{4}{Physics Department, University of California, Santa Barbara, 
CA 93106-9530, USA}
\altaffiltext{5}{University of Nottingham, School of Physics and Astronomy, 
Nottingham, NG7 2RD, UK}
\altaffiltext{6}{Department of Physics and Astronomy, 
Texas A\&M University, 4242 TAMU, College Station, TX 77843-4242, USA}

\begin{abstract}
Stellar feedback  is one of the fundamental mechanisms of galaxy formation and evolution, but observational 
constraints on this process have so far been limited.  In this paper,  
we investigate the properties of feedback-driven shocks in 8 nearby starburst galaxies 
as a function of star formation rate (SFR) and stellar mass, using narrow--band imaging data from the 
Hubble Space Telescope (HST).  The high angular resolution of the HST is a crucial capability to enable 
this type of analysis, as the shock fronts tend to have thin interfaces.
We identify the shock--ionized component via the line diagnostic diagram  
\oiii/\hb vs. \sii (or \nii)/\ha,  applied to resolved regions  3--15~pc in size. 
While we adopt the ``maximum starburst line'' (MSL) from Kewley et al. (2001) to separate shock--ionized 
from photo--ionized gas, we note that the strong metallicity dependence of the  \oiii/\hb ratio requires that we 
divide our sample into three sub--samples:  sub--solar  
(Holmberg II, NGC 1569, NGC 4214, NGC 4449, and NGC 5253), 
solar (He 2-10, NGC 3077) and super--solar  (NGC 5236). This enables us to apply the MSL criterion in an internally--consistent 
manner.  For the sub--solar sub--sample, we derive three scaling relations: 
(1) $L_{shock} \propto {SFR}^{~0.62}$, (2) $L_{shock} \propto {\Sigma_{SFR,HL}}^{~0.92}$ , and 
(3) $L_{shock}/L_{tot} \propto {(L_H/L_{\odot,H})}^{-0.65}$, where $L_{shock}$ is the \has luminosity 
from shock--ionized gas, ${\Sigma_{SFR,HL}}$ the SFR per unit half-light area , $L_{tot}$ the total \has luminosity, 
and $L_H/L_{\odot,H}$ the absolute H-band luminosity from 2MASS normalized to solar luminosity.  
The other two sub--samples do not have enough number statistics,  
but they appear to follow the first scaling relation, i.e. that higher SFRs produce larger  shock \has luminosity.  
The scaling relations indicate that, for stellar feedback: 
(1) when energy is deposited in a small volume, it  produces more shocked \has emission than when deposited in a larger volume; and 
(2) energy deposited in a low mass galaxy produces more shocked \has emission than in a high mass galaxy. 
The energy recovered indicates that the shocks from stellar feedback in our sample galaxies are fully radiative. If 
the scaling relations we derive are applicable in general to stellar feedback, whether it appears in the form of radiative shocks and/or of 
gas outflows, our results are similar to those recently published by Hopkins et al. (2012) for galactic super winds. This similarity should, 
however, be taken with caution at this point, as the underlying physics that enables the transition from radiative shocks to gas outflows  in galaxies 
is still poorly understood.

\end{abstract}
\keywords{galaxies: ISM -- galaxies: interactions -- galaxies: starburst -- ISM: structure}

 \section{Introduction}
 The energy and momentum deposited by star formation activity into the 
 interstellar medium (ISM), a.k.a stellar feedback, is a major, 
 but still not fully characterized, mechanism that governs 
 the formation and evolution of galaxies. 
 The stellar winds and supernovae explosions in star forming regions provide 
 energy and momentum to the surrounding ISM changing 
 its thermodynamic and kinetic properties, and sometimes 
driving galactic scale outflows (Heckman et al. 1990, Martin 1997, Martin et al. 2002, Soto et al. 2012). 
Outflows can eject metals and gas from the host galaxies   
into the intergalactic medium (IGM) enriching the latter in metals and 
suppressing further star forming activity in the galaxies themselves (Oppenheimer \& Dav\'e 2006, Dav\'e et al. 2011).
 Galactic scale outflows/winds also have been called upon 
 to account for the mass-metallicity relation, to shape 
 the luminosity function of galaxies, especially at the faint end slope, 
 and to account for the kinematics of neutral gas 
 in Damped Lyman Alpha systems 
 (Tremonti et al. 2004, Scannapieco et al. 2008, Hong et al. 2010).

Among the poorly constrained parameters from an observational point of view is the 
Òenergy efficiencyÓ of feedback, i.e. the fraction of a starburstÕs 
mechanical energy that is available to drive large-scale outflows. 
Theoretical works (Chevalier \& Clegg 1985; hereafter CC85, 
Silich et al. 2003, Suchkov et al. 1996) suggest that the large supernova (SN) rate 
in a starburst causes the SN remnants to merge together before a 
significant amount of energy is radiated away, and to transfer most of the 
energy outside the starburst volume. 
Such thermalized energy is a main power source for driving superwinds 
from starburst regions.

The hot gas and bipolar winds predicted by the CC85 model are, however,  
too hot to be observed. We mainly observe, at X-ray and optical wavelengths, 
the rapid cooling zones that are most likely associated with the entrainment 
and mass loading of cooler ISM or the boundaries where phases mix 
(Cecil et al. 2002, Martin et al. 2002, Strickland et al 2004). 
These still provide important information: by observing the locations and intensities of the
 outer wind shocks, we can make local estimates of wind energy densities and, 
therefore, constrain the wind power from the starbursts. 

Simulations and models of the propagation of feedback energy on galactic scale 
 (De Young \& Heckman 1994, MacLow \& Ferrara 1999, 
Strickland \& Stevens 2000, Hopkins et al. 2012) indicate that SFR and galaxy mass 
determine the feedback ability to drive galactic outflows, 
though predictions point at galaxies that are about 10-100 times 
less massive than those in which superwinds are observed. Although additional 
complications may arise from the presence of AGNs in massive galaxies, 
these results demonstrate that our understanding of how the mechanical energy from star formation 
interacts with the surrounding gas and how efficient such energy is at driving galactic scale winds 
is still limited. 

In this paper, we investigate the properties of feedback-driven shocks in 8 nearby, AGN--free, starburst galaxies 
as a function of star formation rate (SFR) and stellar mass; 
He 2-10, Holmberg II, NGC 1569, NGC 3077, 
NGC 4214, NGC 4449, NGC 5236 (M83), and NGC 5253. 
There are two main ionizing radiation fields in starburst galaxies: 
the radiation from stars and the cooling radiation from shocks. 
The gas ionized by radiative shocks (shock--ionized gas) generally forms a thin and faint gas layer, while 
the gas ionized by stellar photons (photo--ionized gas) is the dominant  ionized gas component  in starburst galaxies 
and is morphologically diffuse. 
 Because of this characteristic, the luminosity contrast between shock--ionized gas and 
photo-ionized gas is low in low-resolution ground-based observations, and 
the shock--ionized component is thus generally difficult to detect. 
The high angular resolution of the HST, therefore, is necessary for separating shock--ionized gas from  photo--ionized gas in
external galaxies, not only from the line ratios but also in terms of morphology. HST images 
in optical narrow--band filters offer the opportunity to probe both shock--ionized and photo--ionized gas in our 
sample galaxies, over spatial scales in 
the 3--15~pc range, which is small enough to enable separation of the two ionized gas constituents. 
The stringent requirement for high angular resolution is such that 
the results presented in this paper have been mostly unexplored before. 

\section{General Data Analysis Approach}

To identify the emission from shocks, we use the emission line diagnostic diagram: 
\oiii/\hb vs. \sii   (or [N {\footnotesize II}]($\lambda$6583))/\ha). 
The original use of this diagnostic was to discriminate 
starburst from active galactic nuclei (AGN) activity in galaxies, since   
the line ratios are sensitive to the hardness of the 
ionizing radiation field (Baldwin, Phillips, \& Terlevich 1981, 
Kewley et al. 2001; hereafter K01, Kauffmann et al. 2003). 
For this goal, galaxy--averaged line ratios are plotted 
on the diagnostic diagram, each point representing one galaxy. 

As a different application of the same diagram, 
Calzetti et al. (2004; hereafter C04) plotted the line ratios of individual 
regions (bins of 5--10~pc size) from the spatially--resolved images of four nearby starburst galaxies, 
to separate the shock--ionized component from the photo--ionized gas in the galaxies' ISM.  
They adopted the ``maximum starburst line'' (MSL) from K01 as a shock separation criterion; the MSL  
is a theoretical limit,  where the region to the upper--right of the line 
can not be explained by photo--ionization alone. 
C04 find that the estimated \has luminosity from shock--ionized gas 
is a few percent of the total \has luminosity. Their calculations indicate  
 that the mechanical energy from the central starburst 
is enough to power the radiative shocks and a significant amount of the 
mechanical energy is radiated away. 
Hong et al. (2011; hereafter H11) adopt a larger range of properties for the 
photo-- and shock--ionization models 
and apply various shock separation criteria to NGC 5236 (M83). 
The estimated shock \has luminosity can vary from a couple of percent 
(from the ``maximum starburst line''; the most conservative criterion)  
to 30\% (from the most generous criterion \sii/\ha $ > - 0.5$) of the total \has luminosity.  
An intermediate estimate of the \has luminosity in shocks places it at about 15\% of the total, 
which implies that virtually all of the mechanical energy from the starburst is radiated away. 

Building upon the previous studies of C04 and H11, we analyze in this paper  
a larger sample of 8 galaxies and map out the relations between 
the feedback-driven shocks and global galactic parameters: 
SFR ($\sim$ mechanical energy injection rate) and 
host galaxy stellar mass  ($\sim$ gravitational potential depth).
With this, we attempt to specify relations that can inform models of galaxy formation and evolution. 
In \S 3, we describe the sample and data reduction processes. 
In \S 4, we describe the diagnostic diagrams and their related ionization models. 
Then, we present our main results. 
In \S 5, we discuss and summarize our results.

\section{Data Description}

\subsection{Sample Description}

In order to secure a sample of actively star-forming galaxies  
spanning a range in stellar mass and SFR, the following criteria were applied:   
(1) Distance $< 12$ Mpc 
(to exploit the HST angular resolution, $0.2'' =12$ pc at 12 Mpc, 
matched to the observed width of shock fronts in other galaxies; see the \S2 of C04);
 (2) recession velocity $<950$ km/s (to get the emission lines inside the 
 available narrowÐ band filters);  (3) centrally concentrated starÐformation/starburst. 

Observations were performed with the HST through a number of programs (GO - 9144, 10522, 11146, 11360), 
that  gathered narrow-band images 
in F502N, F656N (or F658N, F657N), F487N, and F673N, centered on the relevant emission lines, plus 2 broadÐband 
images in F547M(or F555W, or F550M) and F814W for stellar continuum subtraction. A total of 8 starburst galaxies were observed 
via those GO programs, and a recently approved observing program will secure an additional two (Mrk178 and 
NGC 4861, with program GO-12497). As a general operational approach, 
we produce emission line images for \ha, \hb, \oiii, and \sii (or \nii) 
from the narrow images by subtracting stellar continuum using broad--band images. 

\begin{figure}[t]
\centering
\includegraphics[height=4.0 in]{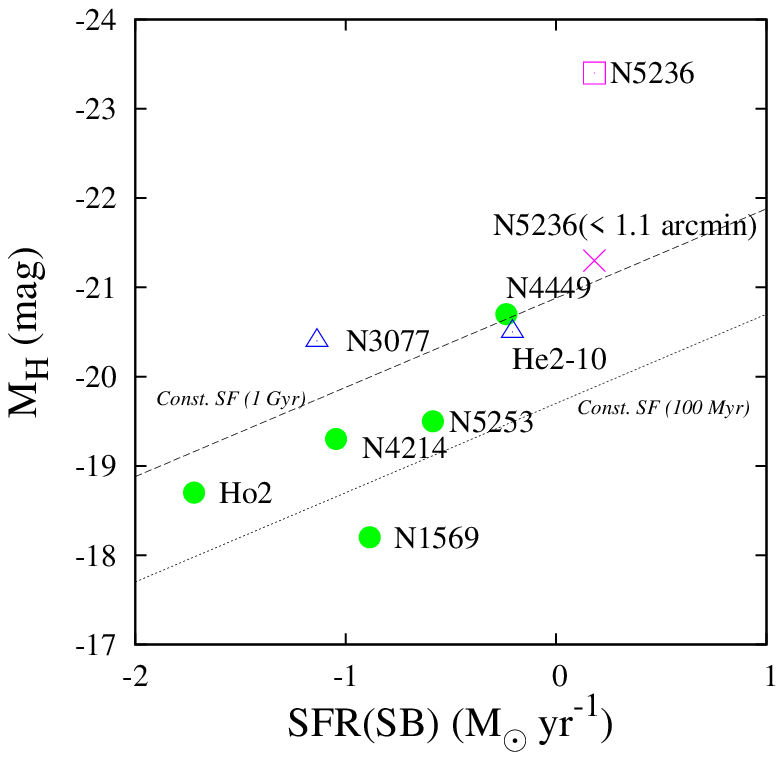}
\caption{ The H-band luminosity from 2MASS  (a proxy of stellar mass; Jarrett et al. 2003) vs. SFR for starburst region (SB), 
shown in Figure~\ref{fig:finalten}, 
for the sample grouped by metal abundance; 
sub-solar (solid circle), solar (open triangle), and super-solar (open square). 
The dotted and dashed lines show the H-band magnitudes derived from STARBURST99 
for constant star forming history with the given period of time; 100 Myr (dot) and 1 Gyr (dash). 
Because NGC 5236 is a large spiral galaxy, 
we also put the H-band magnitude with the aperture size of 1.1 arcmin (Jarrett et al. 2003), 
which is matched to the size of the central starburst region. 
}\label{fig:finalone}
\end{figure}

Figure~\ref{fig:finalone} shows the distribution of our 8 galaxies in 
the parameter space of absolute H--band magnitude from 2MASS (a proxy for stellar mass) vs. 
the SFR within the central starburst region. Table 1 summarizes the general information 
about the sample galaxies and the instrument and filters used in the observations.  
We categorize the sample into three sub--groups 
according to metallicity; sub-solar  
(solid green circles in Figure~\ref{fig:finalone}; galaxies with metallicity around $Z=0.4Z_\odot$.), 
solar (open blue triangles; galaxies with metallicity around $Z=1.0Z_\odot$), 
and super--solar (open purple square; galaxies with metallicity near $Z=2.0Z_\odot$). 
This grouping is guided by the metallicity dependence of the line ratios, which thus affects the 
numerical definition of shocked gas. More details on this effect are given in \S 4.1.

\subsection{Data Reduction}

Our sample was observed using the three HST instruments WFPC2, ACS, and WFC3. 
Table 2 summarizes the narrow--band imaging observations: filter, instrument, target emission line, 
program ID, exposure time, and sensitivity. For ACS and WFC3, we use the standard pipeline image 
products, which include processing through the MultiDrizzle software (Fruchter et al. 2009). 
For WFPC2, the STScI pipeline processing includes only basic steps such as flat-fielding and bias subtraction, 
for which we use the best reference files available in each observing period (Gonzaga \& Biretta 2010). 
Our post-pipeline steps thus include removal of warm, hot pixels and cosmic rays, and registration and 
co--addition of the multiple images in each band.  We use the 
IRAF (Image Reduction and Analysis Facility) task WARMPIX 
to remove hot pixels and the task CRREJ to remove cosmic rays and 
combine the dithered images (see C04 for more details about the WFPC2 post--pipeline processing). 

Then, we perform the four  steps below: 
\begin{enumerate}
\item Photometric calibration. 
\item Stellar continuum subtraction from narrow-band image. 
\item Decontamination of [\iion{N}{ii}]($\lambda\lambda$6548,6583) from narrow-band image for \has if necessary. 
\item Dust extinction correction. 
\end{enumerate}

We use the keyword PHOTFLAM for the photometric calibration of each galaxy in each band.
The filters F487N and F502N only include a single emission line, \hb and \oiii \msp, repectively. This makes 
the photometric calibration simpler than the redder filters. 
For the \sii \msp calibration, we adopt the doublet ratio, [S {\footnotesize II}](6716/6731) $\approx 1.2$. 
This density indicator varies from 1.0 to 1.4 for most ISM conditions and,  
within this range, the calibrated fluxes change by less than 10\%. 
The uncertainty from continuum subtraction is generally larger than this uncertainty. 
For Holmberg II and NGC 4449, we observe the \nii \msp line using the F660N filter, instead of the \sii \msp line. 
The \nii \msp decontamination of the \has filter, thus, is straightforward for these two galaxies. 

The second and third steps produce most of the uncertainty when calculating the line ratios for the diagnostic diagram. 
The fourth step, the dust extinction correction, typically  does not affect the line ratios in the diagnostic diagram in a significant manner, 
due to the proximity in wavelength of the  lines in each ratio, under the assumption that the line emission is coming from the 
same spatial region. However, the dust corrections for each line luminosity can be large. 
The following subsections describe in details the impact of each of the last three steps, and how we deal with them.

\subsubsection{Stellar Continuum Subtraction}

For each narrow-band image, we approximate the stellar continuum baseline 
near the target emission line using broad-band images, straddling, when possible, a line with two adjacent broad--band 
filters. Specifically, we use F547M (or F555W, or F550M) as reference stellar continuum for the \hb and \oiii \msp and an interpolated continuum for \has and \sii (or \nii) using F547M and F814W. For the F555W filter, we remove the self-contamination due to  \hb and \oiii \msp line emission 
within the broad--filter bandpass using the iterative method described in H11. 
To find the optimal subtraction, 
we apply the skewness transition method to all of our narrow-band images (Hong et al. 2013; in prep).  
This method is based on a feature that appears in the skewness at the transition between over-- and under--subtraction of 
the stellar continuum from narrow--band images. The stellar continuum subtraction step is the one that produces the 
largest uncertainty in our final line emission images, due to the intrinsic limitations of the method, which cannot account, e.g., 
for color differences among stellar populations across the filter bandpasses.  

\subsubsection{[\iion{N}{ii}] Correction in the \has Filter}

Except for Holmberg II and NGC 4449, for which the  [\iion{N}{ii}] line was directly observed, assumptions have to be made in 
order to remove the [\iion{N}{ii}] contamination from  the 
narrow--band images targeting the \has emission.  
We use two methods to deal with this problem. 
The first is to use a line ratio [\iion{N}{ii}] / \has obtained from spectroscopy (e.g., from the literature). Because we use a single value for the 
whole image of each galaxy, spatial variation of the [\iion{N}{ii}] / \has ratio can not be considered in this method.  
 The second is to use the relation [\iion{N}{ii}] $ \propto $ [\iion{S}{ii}], which is less dependent on variations in the metal abundance 
and/or UV radiation. With this method, we still assume a constant factor for the entire image. 
Therefore, the  [\iion{N}{ii}] correction is also a limitation for the photometric calibration of the \has images in our sample, although the impact 
of [\iion{N}{ii}]  variations is expected to be significant ($>$10\%) only for M83, which has a large [\iion{N}{ii}] / \has ratio. 
Table 1 summarizes the [\iion{N}{ii}] correction method applied to each galaxy. 

\subsubsection{Dust extinction correction}

We produce extinction maps using the \ha/\hb ratios and the standard extinction equation: 
\begin{equation}
E(B-V) = \frac{\log\big[H\alpha / H\beta \big]_a - \log\big[H\alpha / H\beta \big]_i}{0.4 ( k_{H\beta} - k_{H\alpha})} 
\end{equation}
From Cardelli et al. (1989), we choose a normalization, $R_V = 3.1$, and use the 
Milky Way extinction curve; $k_{\lambda 4861} = 3.609$, $k_{\lambda 5007} = 3.473$, 
$k_{\lambda 6563} = 2.535$, $k_{\lambda 6583} = 2.525$, and $k_{\lambda 6725} = 2.458$, following the recipe of 
Calzetti (2001). This approach enables us to include both internal and foreground (from our own Milky Way) 
extinction simultaneously. 
The dust correction for the line ratios on the diagnostic diagram can be written as :
\begin{eqnarray}
\log \Big( \frac{[S II]}{H\alpha}\Big)_i & = & \log \Big( \frac{[S II]}{H\alpha}\Big)_a - 0.0310 ~ E(B-V) \nonumber \\
\log \Big( \frac{[N II]}{H\alpha}\Big)_i & = & \log \Big( \frac{[N II]}{H\alpha}\Big)_a - 0.0040 ~ E(B-V) \nonumber \\
\log \Big( \frac{[O III]}{H\beta}\Big)_i & = & \log \Big( \frac{[O III]}{H\beta}\Big)_a - 0.0542 ~ E(B-V) \nonumber 
\end{eqnarray}
where the ``i'' subscripts represent the intrinsic line ratio and the ``a" subscripts represent the observed attenuated line ratio.  
The coefficients preceding the color excess E(B-V) are sufficiently small that  dust corrections are in general a minor effect on the 
diagnostic diagram. They are needed only for extreme cases (e.g., for the heavy foreground extinction affecting NGC ~1569). 
However, dust corrections can potentially introduce bias in the case of  diffuse and faint gas, as statistical noise in the  \has  and  \hb maps 
at low surface brightness levels can produce artificially high values of the color excess E(B-V), as will be discussed below.  

\begin{figure}[t]
\centering
\includegraphics[height=5.5 in]{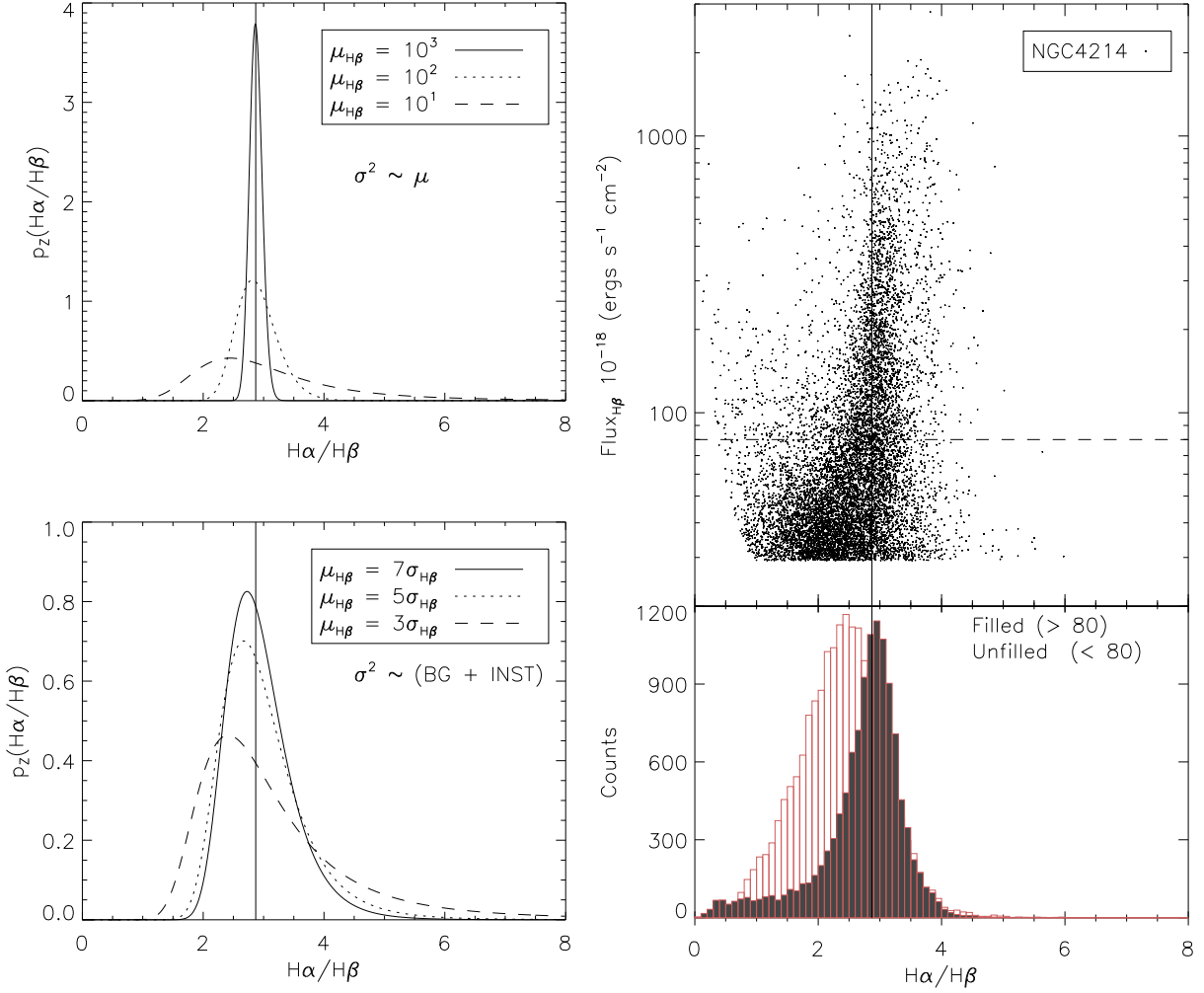}
\caption{ The probability density functions of \ha/\hb ratio for zero extinction, i.e. the intrinsic \ha/\hb = 2.87, 
from Equation 2 for various variances (left panels). 
For a bright source, the noise is dominated by the Poisson fluctuations on the source counts. The left-top panel represents 
this source-dominated case, while the left-bottom panel is for a background and instrument (BG+INST) noise-dominated case.  
The observed \ha/\hb ratio versus the \hb flux for NGC 4214 (right-top panel) and its number count histograms (right-bottom panel). 
The vertical lines represent the positions of \ha/\hb = 2.87. The horizontal dashed line on the right-top panel shows 
the threshold line where the \hb flux is equal to 80 in the given scale unit. 
Above the threshold line (the filled histogram in the right-bottom panel), 
the signal is relatively strong, hence the broadening caused by statistical measurement error is relatively small as 
shown in the left-top panel (the probability function for $\mu_{H\beta} = 10^{3}$).
Below the threshold (the unfilled histogram in the right-bottom panel), 
the statistical broadening becomes larger. Even though we consider other factors affecting the observed ratios, 
such as real dust extinction and residuals from stellar continuum subtraction, the statistical broadening 
shows a significant impact on the observed ratios. This statistical bias can propagate during dust extinction correction. 
}\label{fig:finaltwo}
\end{figure}

The statistical distributions of astronomical data are generally poissonian or gaussian. 
The ratio of two random variables with poissonian or gaussian uncertainty distributions can produce a skewed distribution of 
the ratios. If we assume two normal (gaussian) distributions, expressed as $X = N(\mu_X, \sigma^2_X)$ and $Y = N(\mu_Y, \sigma^2_Y)$, 
where $\mu$ represents the mean and $\sigma$ represents the standard deviation, 
their ratio distribution, $Z \equiv X/Y$, follows the equation below (Hinkley 1969). 
\begin{eqnarray}
p_Z(z) & = &\frac{b(z) c(z)}{a^3(z)} \frac{1}{\sqrt{2 \pi} \sigma_x \sigma_y} \Bigg[ 2 \Phi \Bigg( \frac{b(z)}{a(z)}\Bigg) - 1 \Bigg] \nonumber\\ 
& &+  \frac{1}{a^2(z) \pi \sigma_x \sigma_y} e^{ - \frac{1}{2} \big(\frac{\mu^2_x}{\sigma^2_x} + \frac{\mu^2_y}{\sigma^2_y} \big)} 
\end{eqnarray}
where 
\begin{eqnarray}
a(z) & = & \sqrt{ \frac{z^2}{\sigma^2_x} + \frac{1}{\sigma^2_y} } \nonumber \\
b(z) &=& \frac{\mu_x}{\sigma^2_x} z + \frac{\mu_y}{\sigma^2_y}   \nonumber \\
c(z) &=& e^{ \frac{1}{2} \frac{b^2(z)}{a^2(z)}  - \frac{1}{2} \big(\frac{\mu^2_x}{\sigma^2_x} + \frac{\mu^2_y}{\sigma^2_y} \big)} \nonumber \\
\Phi (z) & = & \int^{z}_{-\infty} \frac{1}{\sqrt{2 \pi}} e^{-\frac{1}{2} u^2} du \nonumber
\end{eqnarray}
Since \has and \hb emissions are strong recombination lines, their intrinsic flux ratio is quite robust in most physical conditions; 
\ha/\hb $\approx 2.87$. Because of this property, Equation 1 is a standard approach to measure the amount extinction from the observed line ratios. 
However, Equation 1 is only valid when our observational sensitivity is infinitely high; in other words, 
the distribution of observed \ha/\hb ratio must follow a delta function, $\delta(z - 2.87)$, when there is no dust extinction. 
Indeed, the ratio distribution (equation~2),  for sufficiently large $\mu_X$ and $\mu_Y$ with poissonian variances ($\sigma^2 \sim \mu$), shows a 
delta function-like distribution (see the probability distribution for $\mu_{H\beta} = 10^3$ in the left-top panel in Figure~\ref{fig:finaltwo}). 
This means that for a bright region the dust extinction correction using Equation 1 is reliable. But, for a faint region, 
the \ha/\hb ratios are heavily affected by statistical measurement errors. 

The important point is 
that, even though there is no dust extinction, the \ha/\hb line ratio can be different from the intrinsic value, 2.87, due to stochastic errors 
and shot noise. This is significant especially for diffuse gas (which is generally faint, implying large stochastic errors). 
The left-bottom panel of Figure~\ref{fig:finaltwo} shows the ratio distributions when the variance is dominated by background (BG) 
or instrumental noise ( dark currents and readout noises; INST), while the left-top panel 
shows the ratio distribution when the variance is dominated by the signal counts and the noise is Poissonian. 
Even for the 7$\sigma$ detections for both cases, the distributions of line ratio are broad. 

The right panel shows the \hb flux versus the observed \ha/\hb ratios   for NGC 4214. Each point is one bin in the images as indicated in
Table~1. NGC 4214 is one of the least dust--extincted galaxies in our sample, as verified from the galaxy--wide \ha/ \hb ratio, 
thus a perfect test case for the present discussion. As can be seen from the right panels of Figure~\ref{fig:finaltwo}, while the 
\ha/\hb ratios spans a small range at high \hb fluxes in the 
pixel--by--pixel\footnote{More correctly, bin-by-bin. To emphasize that our analysis is based on pixel-size scale, not on galactic scale,  
we use the term ``pixel'' for the bins used in this paper.}
distribution (the filled histogram in the right-bottom panel), 
the spread of \ha/\hb ratios increases towards low \hb flux values (the unfilled histogram), 
when we separate the pixels using our selected threshold value 80 ($\approx 14~\sigma$) of \hb flux in the given scale unit. 
The threshold value 80 is high enough, above which the stochastic broadening effect is small. 
While some of the scatter at low flux values may be due to intrinsic variations in the extinction values, 
we cannot exclude that a portion may be due to the statistical fluctuations presented in this section. 

\begin{figure}[t]
\centering
\includegraphics[height=5.5 in]{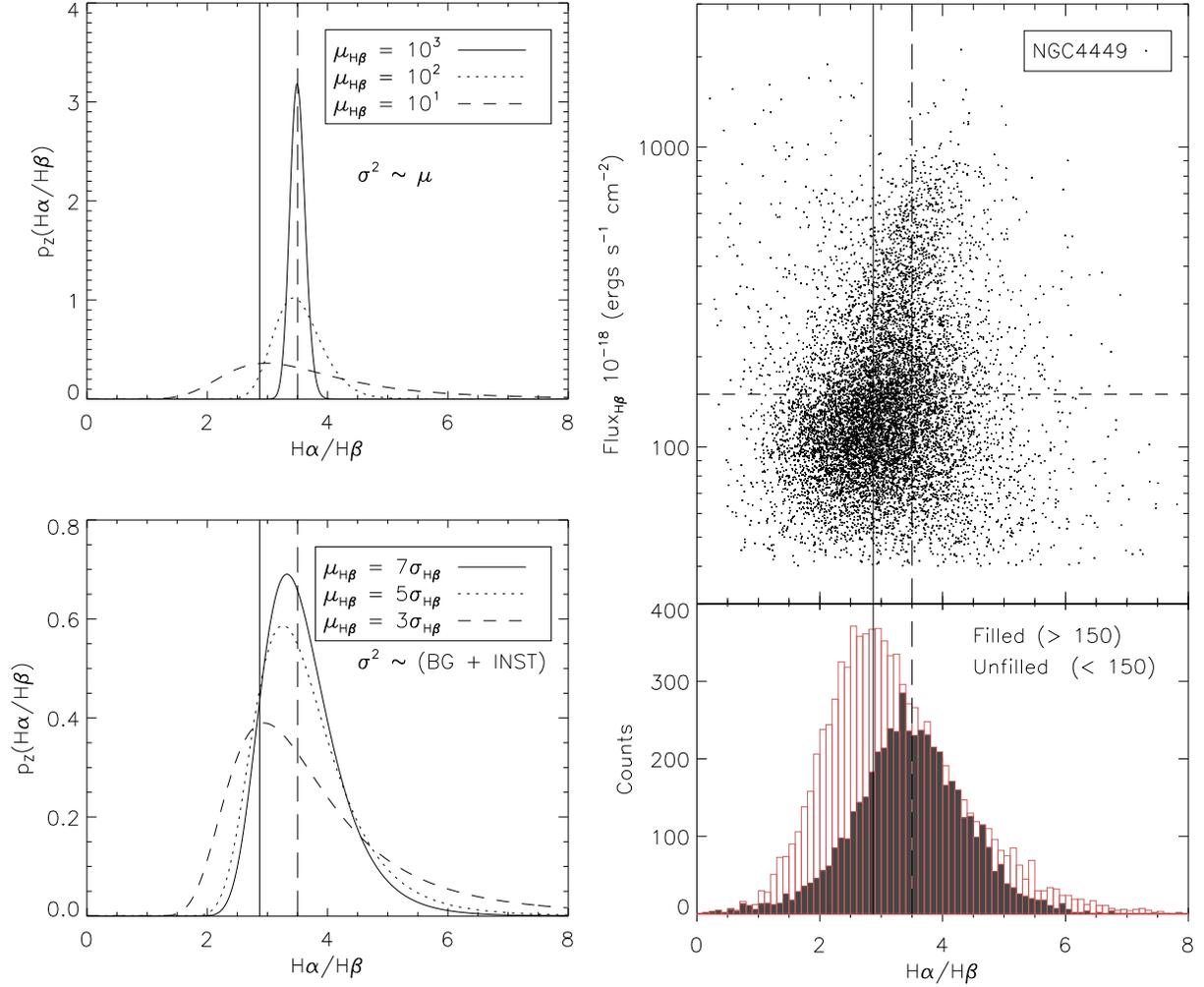}
\caption{ The same figure with Figure~\ref{fig:finaltwo} for NGC 4449. Here we set the intrinsic ratio as \ha/\hb=3.5 
corresponding to E(B$-$V) = 0.2, which is the centroid of the filled histogram in the right-bottom panel. 
Due to the different depth of observation for NGC 4449, 
we set the threshold line at 150 in the given scale unit. The solid vertical lines represent the ratio, \ha/\hb=2.87, the same 
with Figure~\ref{fig:finaltwo} and the long-dashed vertical lines the ratio, \ha/\hb=3.5. The probability functions (left panels) look 
similar with the ones in Figure~\ref{fig:finaltwo}, but they become broader. When we average out the histogram for dust extinction, 
the statistical bias can be alleviated. Therefore, galaxy-averaged correction can suffer less from this bias. 
But for pixel-by-pixel correction, especially that E(B$-$V) is set to zero when \ha/\hb $< 2.87$, 
the corrected line flux is typically overestimated due to this statistical broadening issue. 
}\label{fig:finalthree}
\end{figure}

To show a possible bias due to the statistical fluctuations described above, we present  the case of NGC 4449 in Figure~\ref{fig:finalthree}; 
this galaxy has a non-negligible amount of internal dust extinction. The left panels show the probability functions 
for \ha/\hb = 3.5 corresponding to E(B$-$V) = 0.2, which is the centroid of the filled histogram of the right-bottom panel. 
In the left panels, the overall distributions look similar with the ones in Figure~\ref{fig:finaltwo}, but they become broader. 
The right panels show the observed ratios. 
Due to the shallower depth of the observations for NGC 4449, we set the \hb threshold line at 150 ($\approx 19~\sigma$) in the given scale unit. 
Because of the internal dust extinction, the observed ratios come from the convolution of the 
dust extinction distribution with its statistical broadening. For the brighter pixels above the threshold 
(the filled histogram in the right-bottom panel), the `true' mean value is recovered, despite the broadening due to 
the statistical fluctuations. For the pixels below the threshold, the statistical fluctuations dominate the observed line ratios, and 
the peak of the distribution is below the `true' mean value. 
It is inevitable, for pixel--by--pixel correction, that we have negative extinctions when \ha/\hb $< 2.87$. 
We set E(B$-$V) to zero for those pixels. For galaxy-averaged measurements, the ratios can be averaged out 
by summing all of the pixels. But for pixel--by--pixel correction, the step of imposing E(B$-$V)=0 for otherwise negative extinction values  
causes overcorrection for dust extinction effects. 

To summarize, the issue described in this section is present for pixel--by--pixel analyses of low surface brightness 
(high statistical uncertainty) regions; it is a minor effect in high surface brightness regions or in galaxy--averaged quantities. 
In light of the above, as a general rule, we will use the line ratios in the diagnostic diagrams without extinction corrections. However, 
such corrections will be important for line fluxes, and they will be applied. We will need to keep in mind, however, the caveats discussed in this 
section.

\subsubsection{Error bars and most uncertain ratios (MURs) on diagnostic diagram}

We impose a minimum threshold value of 5~$\sigma$ to our emission line images, in order to construct line ratios (see section 4.4.2 for 
a discussion on changing the threshold). This implies that the uncertainty on the emission line ratios will be the largest 
when both emission lines are detected at the threshold value; any other line ratio will have lower uncertainty values, but will 
also come from brighter line value(s). The ratio involving two emission lines detected at the threshold value will thus 
determine a  pivot ratio, with two properties: 
(1) the deepest detection (i.e., lowest flux value) and (2) the largest error bar. 
Hereafter, we call this ratio as Most Uncertain Ratio (MUR). 

We locate the position of the MUR in each diagnostic diagram, along both axes. We call the intersection of the 
two MUR values the ``MUR spot''.  By adjusting the detection threshold or the observational depth, we can move the MUR spot. 
For example, if all images are observed with equal depth (in flux), the MUR spot is positioned at $(0,0)$ in log scale. If, conversely,  
\sii \msp and \oiii \msp are observed with ten times the depth of the \has and \hb lines, respectively, the MUR spot moves to $(-1,-1)$. 
This adjustment can be useful if we have a specific target ratio: for example, for shocks $\log($ \sii/\ha$) > 0.0$.  
Figure~\ref{fig:finalfive} and ~\ref{fig:finalsix} show the MUR spots with their error bars in our diagnostic diagrams; 
the error bars are FWHMs of ratio distributions. The thin (cyan) error bars are for $5\sigma$ detections, while  
the thick (blue) error bars are for $10\sigma$ detections, hence the latter are smaller than the former. 
Any other line ratio that is not on the MUR spot will have smaller error bars than those at MUR spot.

\section{Results}

\subsection{Diagnostic diagram and metallicity dependence}

\begin{figure}[t]
\centering
\includegraphics[height= 3.5 in]{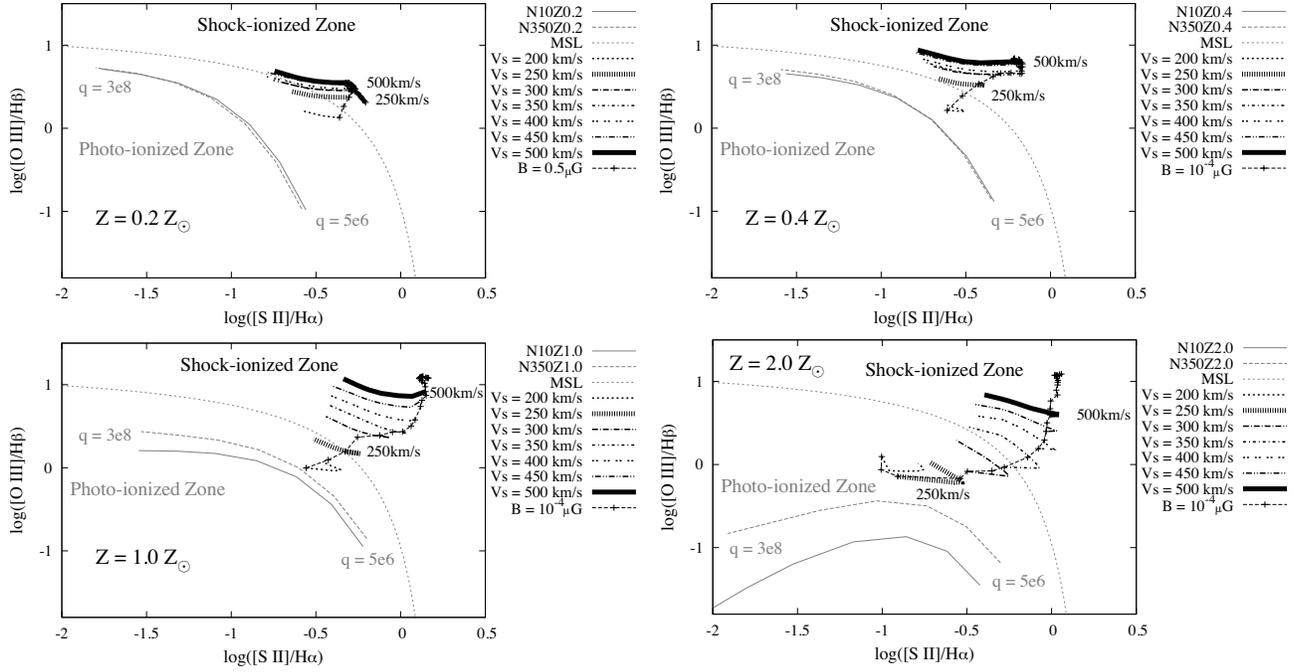}
\caption{ Tracks of the photo-ionization models (Kewley et al. 2001) and the shock--ionization models (Allen et al. 2008) 
for the metallicities, $0.2 Z_\odot$, $0.4 Z_\odot$, $1.0 Z_\odot$, and $2.0 Z_\odot$. 
The MSL is the ``maximum starburst line'' in Kewley et al. (2001).  
The tracks' location depends on metallicity, especially for the photoionized gas. 
N10Z0.2 represents the photoionization model for the ISM density, 10 $cm^{-3}$, and the metallicity, $Z = 0.2 Z_\odot$; 
and the other photoionization models are named in the same way with N10Z0.2. 
For shock models, the lines of $B = 0.5 \mu G$ and $10^{-4} \mu G$ are the main 
branches showing the velocity dependence of shock--ionization grids. 
The side branches of given shock velocity $V_s$ show the magnetic field effects. 
We use $B = 0.5 \mu G$ models for $Z = 0.2 Z_\odot$ instead of $B = 10^{-4} \mu G$, 
because the solutions do not converge in low metallicity and magnetic field. 
The density of preshock gas is 1 $cm^{-3}$. 
}\label{fig:finalfour}
\end{figure}

\begin{figure}[t]
\centering
\includegraphics[height=7.5 in]{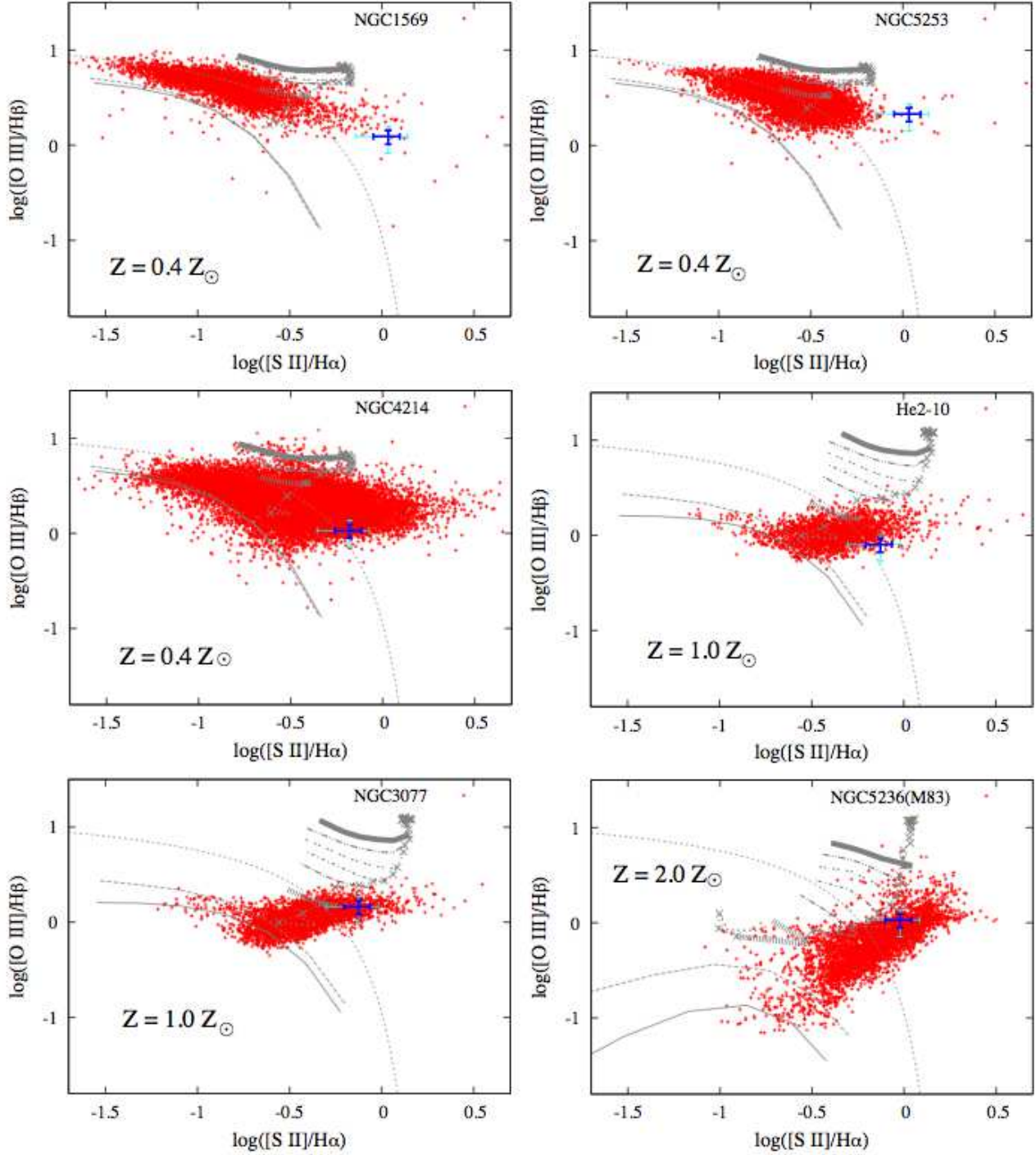}
\caption{ The [\iion{S}{ii}] diagnostic diagram for each galaxy with theoretical grids.
The grey lines show the theoretical grids presented in Figure~\ref{fig:finalfour}. 
The error bars are presented at MUR spots; thin (cyan) error bars for $5\sigma$ detections 
and thick (blue) error bars for $10\sigma$ detections. 
}\label{fig:finalfive}
\end{figure}

\begin{figure}[t]
\centering
\includegraphics[height=8.0 in]{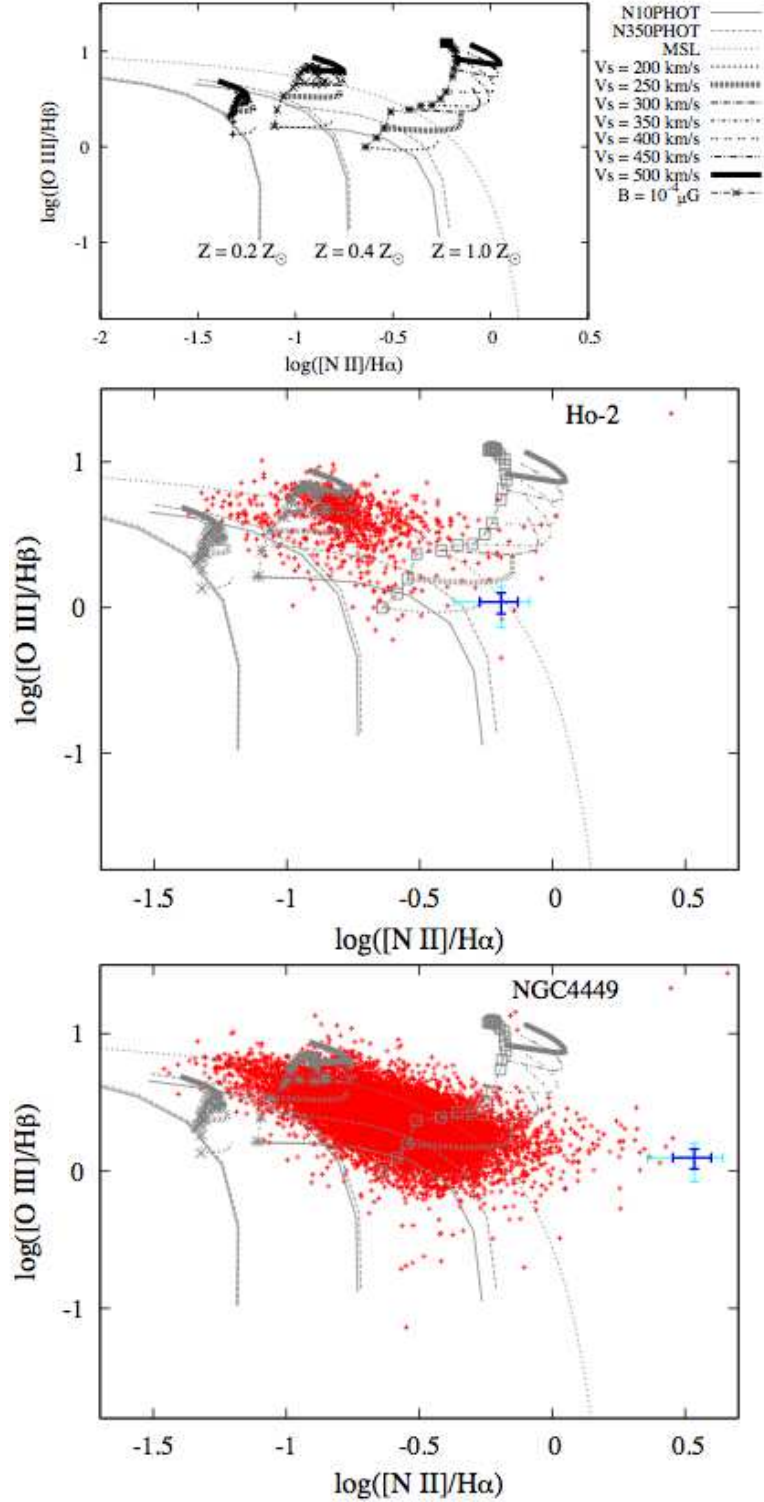}
\caption{Theoretical grids for the [\iion{N}{ii}] diagnostics like Figure~\ref{fig:finalfour} (top). We plot all the models 
with the three metallicities, $Z = 0.2, 0.4, 1.0 Z_\odot$, in this single panel. The same error bars as presented in Figure~\ref{fig:finalfive}. 
Diagnostic diagrams for Holmberg II (middle) and NGC 4449 (bottom) with theoretical tracks presented on the top panel. 
The [\iion{N}{ii}] diagnostics shows the clear separation between different metallicities, but more overlaps between 
shock-- and photo--ionization grids than the [\iion{S}{ii}] diagnostics. 
}\label{fig:finalsix}
\end{figure}

In order to establish whether the line emission from  each of the sub--arcsecond bins in our galaxies is dominated by photo--ionized gas or 
shock--ionized gas, we need to compare the observed line ratios against models.  For this purpose, we adopt the theoretical grids of K01 
for photoionization and Allen et al. (2008; hereafter A08) for radiative shocks. 
The photoionization grids from K01 are based on the stellar population synthesis model STARBURST99 and the gas ionization 
code MAPPINGS III (Binette et al. 1985; Sutherland \& Dopita 1993; Leitherer et al. 1999). 
The spectral energy distributions (SEDs) from STARBURST99 provide the ionizing fluxes, 
and the MAPPINGS III code calculates the ionization state for the atomic species and the fluxes of the emission lines. 
We derive line ratios from those model fluxes for a range of ionization parameter values ($q$, 
defined as the ratio of mean ionizing photon density to mean atom density in K01, ranges from $5.0\times 10^{6}$ to $3.0\times 10^{8}$), and for selected values of the metallicity and density of the gas. 
Figure~\ref{fig:finalfour} shows the photo--ionization tracks for a constant star formation history, Geneva stellar evolution tracks (Schaller et al. 1992) 
and Lejeune stellar atmosphere models (Lejeune et al. 1997). We only need these tracks to provide a consistent way to separate photo-- 
from shock--ionized gas, and not for quantitative analysis.  
Thus, we adopt the conservative track termed ÒMaximum Starburst LineÓ (MSL in the legend of Figure~\ref{fig:finalfour}) of K01 as our separating 
criterion between the two gas components. Above and to the right of this track, line ratios cannot be explained by photons from synthesized stellar 
populations. In our case, we consider this non--stellar ionization to be shocks generated by stellar  mechanical feedback. 

The shock--ionization models of A08 calculate the ionizing radiation field from hot 
radiative shock layers dominated by free--free emission and use the MAPPINGS III code 
for the gas ionization state and the intensity of the emission lines. 
For radiative shocks, the emission comes from two components: the shock layer (post-shock component) 
and the precursor (pre-shock component). The shock layer is the cooling zone of the radiative shock, 
and the precursor is the ionized region by upstreaming photons from the cooling zone. 
Since the main radiation process in the shock layer is free--free emission, 
the ionizing radiation field from shocks is mainly determined by the shock Mach number (i.e., shock velocity), 
the pre--shock gas density, and the intensity of the ISM magnetic field (see A08 for more details). The ISM magnetic field 
affects the post--shock gas density; higher ISM magnetic fields result in lower post--shock densities which affect  
the ionization parameter of the post-shock  gas component. 
For a given metallicity and pre-shock gas density,  the line ratios are thus determined mainly by the shock velocity and,
as a second main parameter, by the magnetic field. 
This is shown in Figure~\ref{fig:finalfour},  for shock velocities from 200 km/s to 500 km/s, 
with a minimum ISM magnetic field of $10^{-4} \mu G$ or $0.5 \mu G$. 
The side branches from the selected shock velocities, 
200, 250, . . ., 500 km/s, show the effect of changing the magnetic field strength from $10^{-4} \mu G$ 
(or $0.5 \mu G$) to $10 \mu G$.  
This discussion assumes that the ambient photoionization rates are small. If that is not the case and 
the cooling zone is photoionized or if projection effects are important, some fraction of the shocked gas may be missed because of dilution. 

Given the emission lines available for our  sample galaxies, we use two diagnostics, 
\oiii /\hb vs. \sii /\ha (hereafter, [\iion{S}{ii}] diagnostics) 
and \oiii/\hb vs. \nii /\ha (hereafter, [\iion{N}{ii}] diagnostics), as summarized in Table 1. 
Since the \oiii /\hb and \nii /\has ratios strongly depend on metallicity (Kewley \& Dopita 2002), 
the shock identification through the MSL is inevitably affected by metallicity. 
Figure~\ref{fig:finalfour} shows the theoretical tracks for the [\iion{S}{ii}] diagnostics 
at four different metallicities,  $0.2 Z_\odot$, $0.4 Z_\odot$, $1.0 Z_\odot$, and $2.0 Z_\odot$. 
The grids show the metallicity dependence of the \oiii / \hb ratio, especially conspicuous for $Z \geq 1.0 Z_\odot$. 
The shock tracks also show different patterns for each metallicity. 
This is easily seen by choosing a constant shock velocity, e.g., v=250 km/s, which is a typical velocity for the observed shocks (see below), and 
follow the variations of this track for changing metallicity. The most noticeable effect in Figure~\ref{fig:finalfour} 
is a decrease of the expected  \oiii / \hb shock line ratio for increasing metallicity.  
The top panel in Figure~\ref{fig:finalsix} shows the theoretical tracks for the [\iion{N}{ii}] diagnostic, which  also presents a clear metallicity 
dependence of the \nii / \has ratios. 

Figure~\ref{fig:finalfive} shows the six galaxies for which the [\iion{S}{ii}] diagnostic is used. 
From the distribution of pixels on the diagnostic diagram, 
we can group the galaxies as (NGC 1569, NGC 5253, NGC 4214), (He 2-10, NGC 3077), and (NGC 5236). 
This grouping corresponds to similar oxygen abundance within each group (Table 1).
An important consideration is that the pixel distribution of each galaxy roughly follows the expected trend at the galaxy's metallicity 
value, as shown by the over-plotted theoretical tracks at metallicities $0.4 Z_\odot$, 
$1.0 Z_\odot$, and $2.0 Z_\odot$.  In addition, the fraction of pixels assigned to shock--ionization is highly dependent on the 
galaxy metal content. More metal rich galaxies will tend to have their shock--ionized gas fraction underestimated, because their 
photoionization track is lower than that of a metal--poor galaxy. 
Because of the metallicity dependence of the line ratios in the diagnostic diagram, we maintain a 
metallicity grouping for the rest of our analysis. Specifically, we divide our sample into  a sub--solar group, a solar group, and a super--solar group, 
as given by the galaxies metallicity. Within each constant metallicity group, the data (e.g., shock identification, energy content, etc.) 
can be compared in an internally--consistent manner.  

For the [\iion{N}{ii}] diagnostics, we only have two sub-solar galaxies in terms of metal abundance, Holmberg II and NGC 4449. 
Tentatively, we put them into the sub-solar group. And, because the MSL is a theoretical limit for photo-ionization regardless of ionic species, 
we assume that the MSL for [\iion{N}{ii}] diagnostics is equivalent to the MSL of [\iion{S}{ii}] diagnostics for shock estimates.

\subsection{[N \footnotesize{II}] versus [S \footnotesize{II}] diagnostics}

Figure~\ref{fig:finalsix} shows the theoretical tracks for the [\iion{N}{ii}] diagnostic (top) 
and the pixel diagnostic diagram for Holmberg II (middle) and NGC 4449 (bottom). 
The main difference between the [\iion{N}{ii}] diagnostic and the [\iion{S}{ii}] diagnostic
is the amount of overlap between shock--ionization tracks and photo--ionization tracks.  
When we compare the two diagrams, 
we find that the [\iion{S}{ii}] diagnostic has smaller overlap between photo-ionization tracks and 
shock--ionization tracks than the [\iion{N}{ii}] diagnostic. 
The better shock discrimination offered by the [\iion{S}{ii}] diagnostic makes this diagram a preferred choice for 
this type of analysis.

However, the shock tracks of the [\iion{S}{ii}] diagnostic at different metallicities  overlap considerably, while 
 the [\iion{N}{ii}] diagnostic shows a better separation among the tracks. 
To summarize, the [\iion{S}{ii}] diagnostic is effective at separating shocks from ionized gas and 
the [\iion{N}{ii}] diagnostics is effective at separating gas components at different metallicity.

\subsection{Single line ratio diagnostics}
\begin{figure}[t]
\centering
\includegraphics[height=8.0 in]{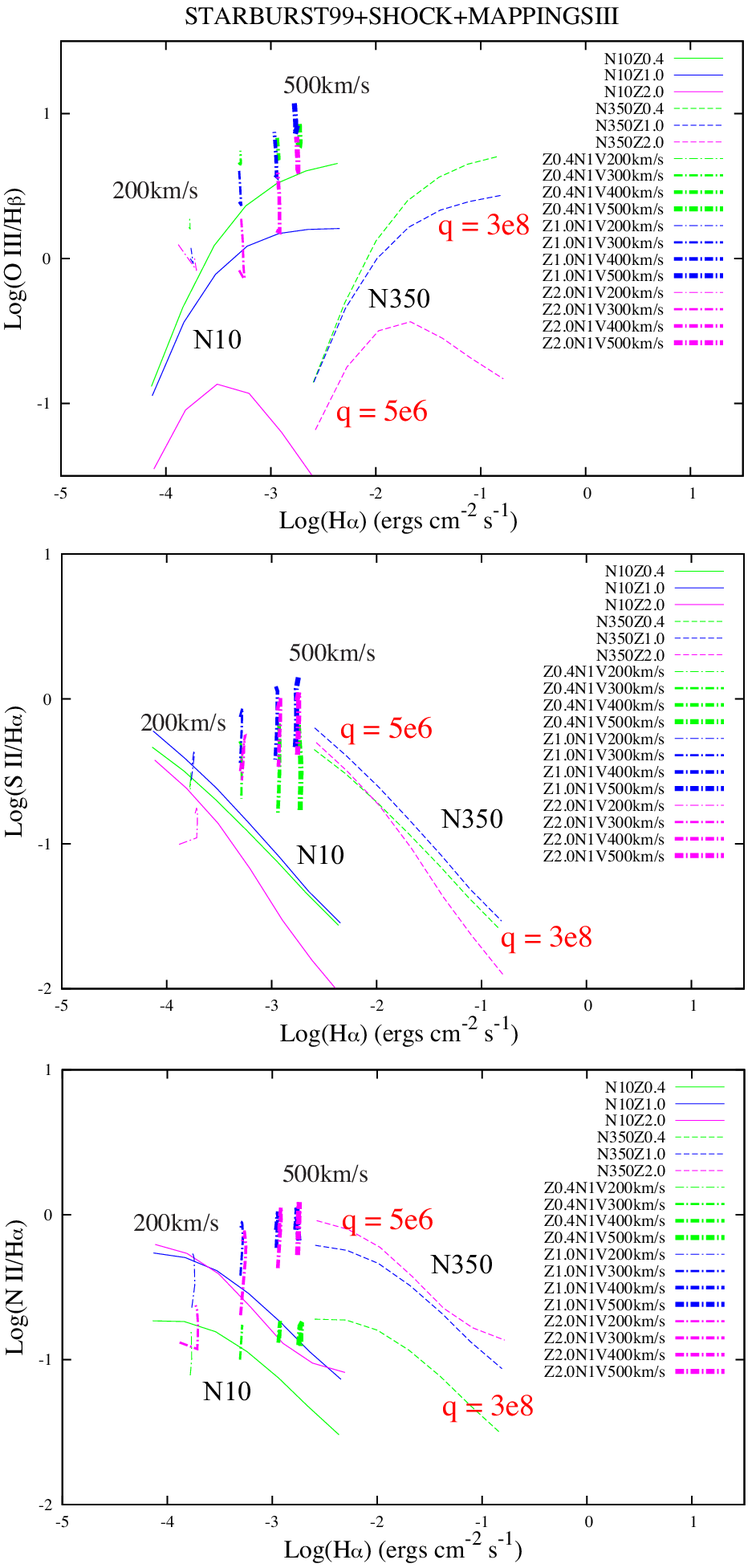}
\caption{ The line ratio of \oiii/\hb (\sii/\ha, \nii/\ha) as a function of the normalized \has from K01 and A08. 
The models are the same ones presented in Figure~\ref{fig:finalfour} and Figure~\ref{fig:finalsix}. 
We plot the grids with different colors for each metallicity; $Z = 0.4 Z_\odot$ in green, $Z = 1.0 Z_\odot$ in blue, and $Z = 2.0 Z_\odot$ in magenta. 
The N350 photoionization models 
are located on the right-hand side of the N10 models due to their higher density; so stronger \has emission.  
The shock--ionized gas has generally higher line ratios shown as vertical lines in the plots for all of the three line ratios than photo--ionized gas. 
This property can be used as a shock separator, using $\log$(\sii/\ha$) > - 0.5$. 
We can also find that the \oiii/\hb ratios in \has bright regions show a strong metallicity dependence as previously shown in Figure~\ref{fig:finalfour}.  
}\label{fig:finalseven}
\end{figure}

When only a single line ratio is available due to limited observational conditions, 
this partial information still can be used for obtaining a rough measure of the  metallicity or an approximate 
separation between shocks and photo--ionized  components.  The derived values will be less reliable than in the 
two--line ratios disgnostics, due to degeneracies of the physical quantities in the single--line ratio case. 
Figure~\ref{fig:finalseven} shows the theoretical tracks for each line ratio versus the normalized \has flux. 
For the \oiii/\hb ratio, the shock tracks are well separated from the photoionization tracks for $Z = 2 Z_\odot$, but 
they tend to overlap for lower metallicity values ($Z \leq 1.0 Z_\odot$); rather, 
the ratio can be used as a metallicity indicator. 

The \sii/\has ratio can be used to separate shocks from photoionized gas, e.g., by employing boundaries such as 
 $\log$(\sii/\ha) $> -0.5$ for shocks. 
The \nii/\has ratio has a similar trend as the \sii/\has ratio, but shows a larger degree of degeneracy than the \sii/\has ratio, 
implying that it is a less effective single--line shock identifier. 

We should note that large [\iion{N}{ii}]/H$\alpha$ and [\iion{S}{ii}]/H$\alpha$ ratios can be obtained also in the presence of hot, diffuse 
photoionized gas (Reynolds et al. 1999 and 2001). This gas will be characterized by weak [\iion{O}{iii}]/H$\beta$ ratios. In the absence of this line ratio, hot 
photoionized gas can be discriminated from shock--ionized gas from the morphological differences: the photoionized gas 
tends to be more uniformly distributed, while the shock--ionized gas is present in thin, shell--like regions. Overall, the use 
of a single line ratio as a shock diagnostic will produce an overestimate in the amount of shocks present in a region. 

\begin{figure}[t]
\centering
\includegraphics[height=8.0 in]{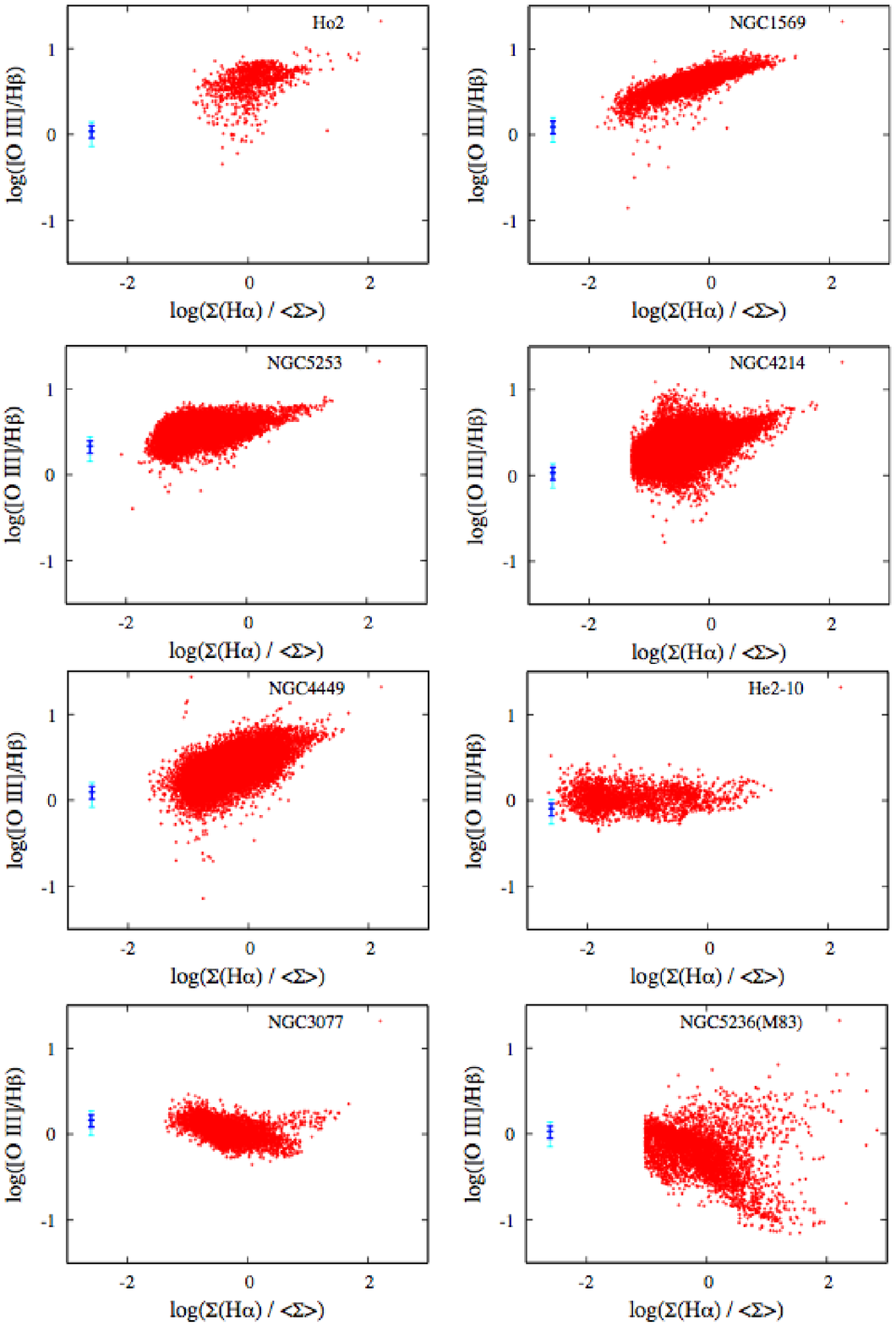}
\caption{The line ratio of \oiii/\hb as a function of the normalized \has surface brightness with 
the error bars presented in Figure~\ref{fig:finalfive}.
The surface brightness is normalized to the half-light radius surface brightness $<\Sigma>$; 
i.e. the sum of all pixels over this value is equal to the half of the total \has flux. 
The metallicity dependence of \oiii/\b ratios in the bright end of \has flux fits well with 
the theoretical grids shown in the top panel of Figure~\ref{fig:finalseven}. 
We can recognize the three groups from the figures; 
the sub-solar group(NGC 1569, NGC 5253, NGC 4449, Holmberg II, NGC 4214), 
the solar group (He 2-10, NGC 3077), and the super-solar groups (NGC 5236). 
Due to the dust lanes in NGC 5236, the horizontal pixel distribution near \oiii/\hb $\approx 0.0$ 
is a propagated error from dust extinction correction. 
}\label{fig:finaleight}
\end{figure}

\begin{figure}[t]
\centering
\includegraphics[height=8.0 in]{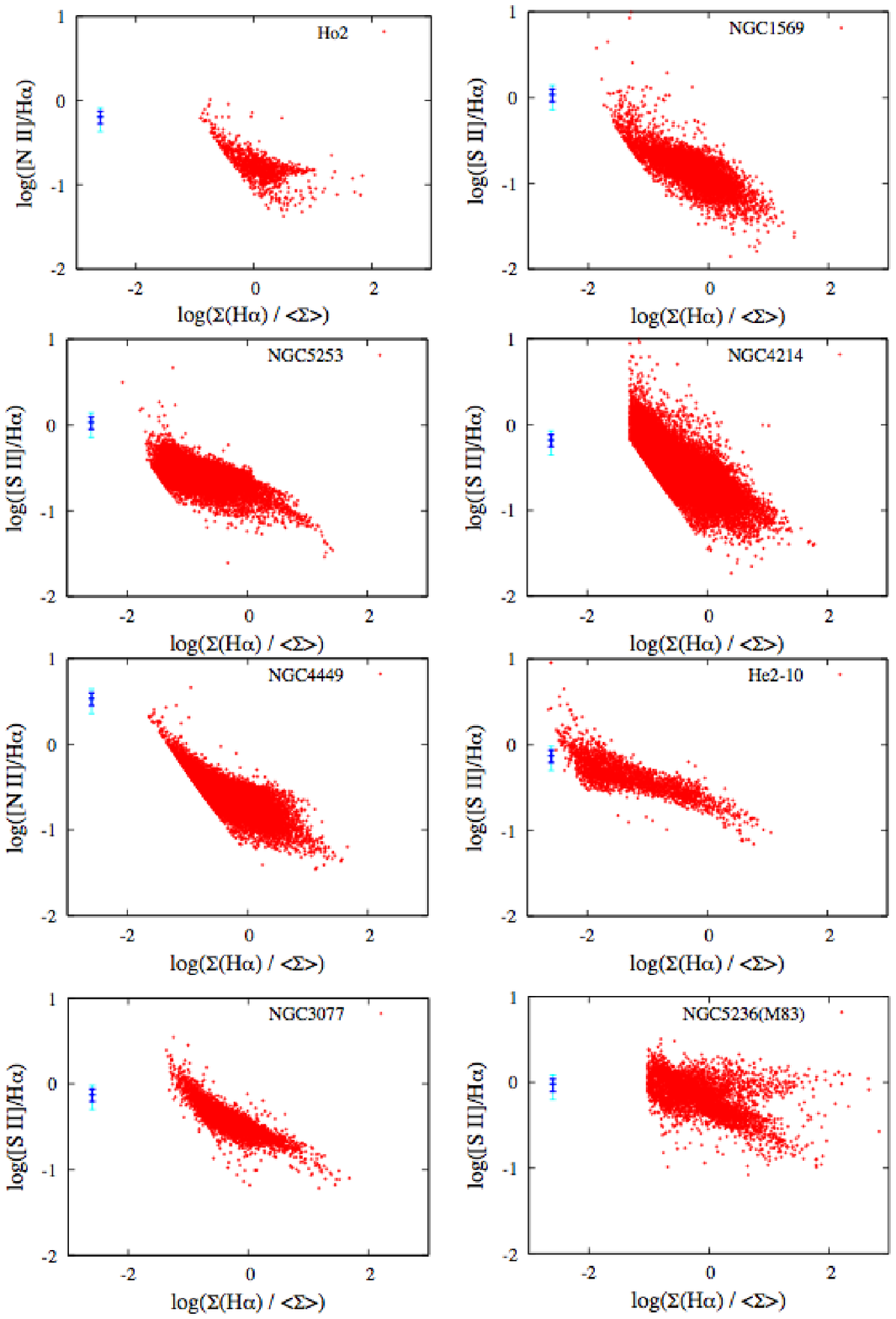}
\caption{The line ratio of \sii/\has (and \nii/\has for Holmberg II and NGC 4449) 
as a function of the normalized \has surface brightness with the error bars presented in Figure~\ref{fig:finalfive}.
We can use this ratio for shock separation, such as 
$\log ($\sii/\ha $) > -0.5$, though the estimated shock is less reliable than the estimates from the diagnostic diagrams, because 
of the potential contamination from hot, diffuse photo--ionized gas.  
}\label{fig:finalnine}
\end{figure}

Figure~\ref{fig:finaleight} shows the observed pixel line ratio \oiii/\hb as a function
of normalized \has surface brightness for our galaxies. 
The metallicity dependence of the \oiii/\hb ratio at the bright end of the \has flux follows theoretical 
expectations  (Figure~\ref{fig:finalseven}). 
We can recognize the three metallicity groups again from the \oiii/\hb trends:  
the sub--solar (NGC 1569, NGC 5253, NGC 4449, Holmberg II, NGC 4214), 
solar (He 2-10, NGC 3077), and super--solar  (NGC 5236). 
Figure~\ref{fig:finalnine} shows the observed line ratio \sii / \ha 
(and \nii/\has for Holmberg II and NGC 4449) 
as a function of the normalized \has surface brightness. Again, 
the observed line ratios follow theoretical expectations. 
Because of the higher sensitivity of the HST instruments at red wavelengths, these single--line ratios can be obtained for a larger 
number of pixels than in the case of two--line ratios, as the line images are deeper in the red than in the blue, 
but provide a poorer separating shock--photo ionization power than the diagnostic diagram. 
Internal variations of the gas density and ionization parameter can additionally cause the scatter of the data points (Figure~\ref{fig:finalseven}). 
Finally, the narrower distribution for higher \has brightness can be due to the property of 
the ratio distributions, which are narrower for higher SNR detections.

\subsection{Relations between Shock--ionized Gas and Stellar Feedback}

\begin{figure}[t]
\centering
\includegraphics[height=7.5 in]{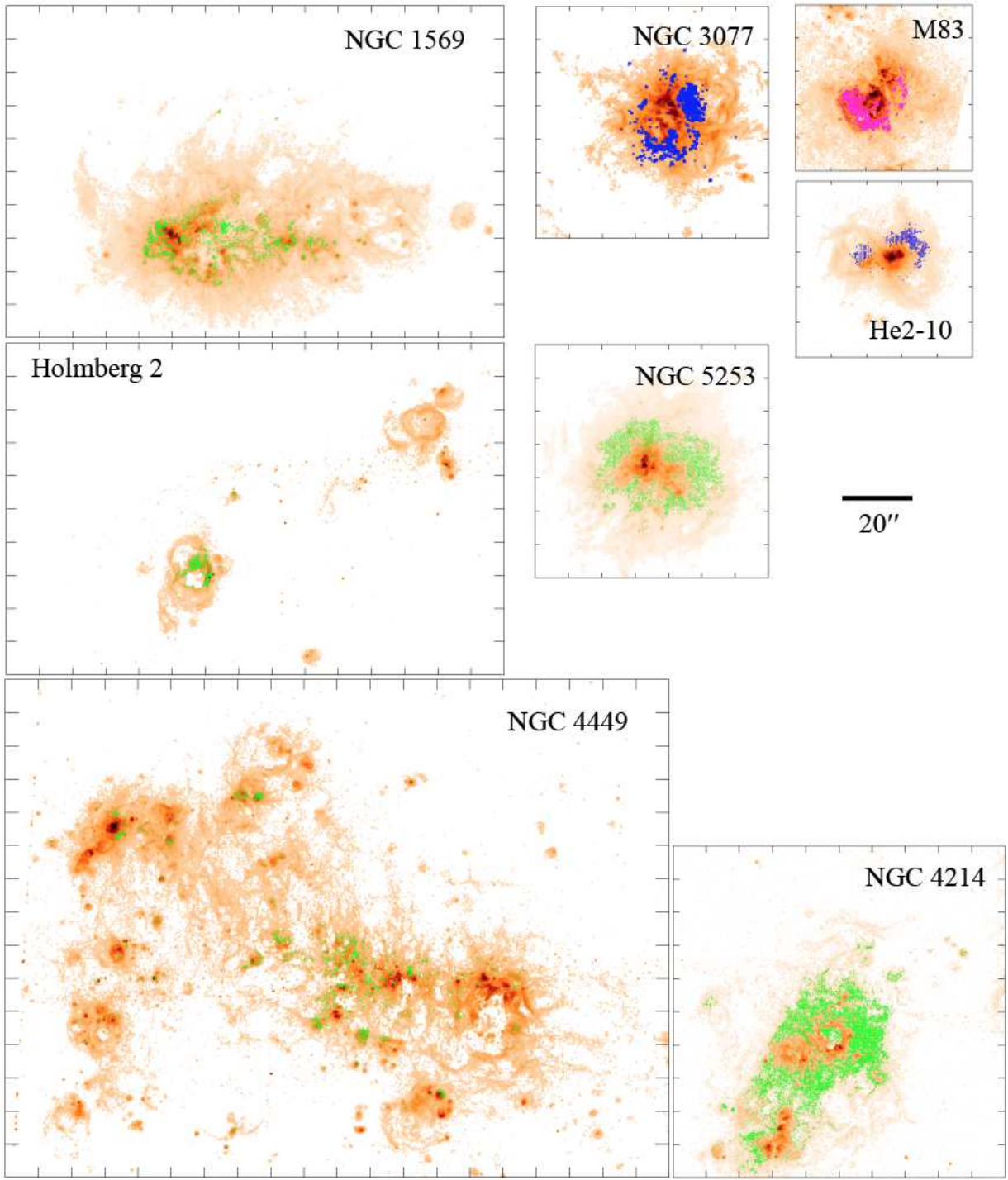}
\caption{Distribution of identified shocks over the \has images 
(painted in green for sub-solar sample, blue for solar sample, and magenta for super-solar sample).  
Most identified shocks are located in the outer rim from the central or most prominent starburst. 
}\label{fig:finalten}
\end{figure}

In this section, we will present our main results on the shock--ionized gas 
and its relations with the H-band magnitude (a proxy of stellar mass) 
and star formation (and star formation density). 
Using the MSL within each metallicity sub--sample, we separate pixels dominated by the shock ionization  
from those dominated by photo--ionization on the diagnostic diagram. 
Figure~\ref{fig:finalten} shows the physical location of the identified shocks, re--projected onto the \has 
map of each galaxy. The pixels dominated by shock--ionization are generally scattered 
in the outer rim of the central starburst.  

From the identified shock--dominated pixels, we measure the two quantities. 
The first is the \has luminosity of the shocked component. 
If the shock is fully radiative, the mechanical luminosity driving the shock 
is about 20 to 80 times larger than the shocked \has luminosity (Rich et al. 2010). 
The numbers are dependent on the specific model, but generally we can adopt that 
a couple of percent of the total shock luminosity is radiated away through the \has emission. 
Hence, by measuring the \has luminosity of the shocks, 
we can estimate the underlying mechanical luminosity which is radiated away.  
The second is the ratio of the \has luminosity of the shocked component to 
the total \has luminosity. The total \has luminosity is linked to the current SFR of the starburst. 
By adopting some appropriate history of star formation, we can 
calculate the mechanical luminosity from the SFR. 
Therefore, the ratio between the shocked and total \has luminosity is an indicator of the balance 
between the energy injection rate and the energy loss rate.  

When measuring the shocked \has luminosity, we have three major factors affecting such a luminosity. 
The first one, already discussed in detail in previous sections, is the shock/photo--ionization separation method, which is 
affected by a strong degeneracy linked to the presence of  a continuum between the two gas components, rather than 
an abrupt transition. 
Within this framework, we suggest that the most important aspect for shock separation is consistency, rather than accuracy. 
When we apply the MSL to our sample, the estimated shocks are consistent within each metallicity sub--sample  
due to the similar distribution on the diagnostic diagram; i.e. similar cooling pattern due to similar metallicity. 

The second one is a threshold driven by the  observational detection limit. 
We adopt a $5\sigma$ threshold for each line emission. This is, however, an artificial threshold imposed by the depth of each image.
In section~4.4.2, we will apply a physical threshold for better consistency among different galaxies, although   
it will turn out that the qualitative results do not change if an artificial or physical threshold is applied. 

The third one is dust correction.
As previously discussed, the dust correction on a pixel-by-pixel basis 
can suffer from bias due to stochastic uncertainty. 
But for heavily dust obscured galaxies, the correction is necessary though 
the error can propagate to the shock estimates. In our sub--solar sample, only NGC 1569 and NGC 4449 have some changes 
from this correction. Like the detection threshold issue, the dust correction does not change the qualitative results. 

Finally, two additional sources of confusion can lead one to the underestimate and the other to the overestimate of the shock 
luminosity. The first is the presence of shocks in strongly photoionized regions; in this case, the shock emission will be diluted, and 
we will tend to underestimate the shock luminosity. The second is the presence of hot, diffuse photo--ionized gas; in this case, 
the low--ionization lines will be enhanced, 
and, especially if utilizing a single line ratio diagnostic (e.g., [\iion{S}{ii}]/H$\alpha$), the diffuse photo--ionized gas may be mistaken 
for shock emission and the shock luminosity overestimated. The latter scenario can, however, be controlled by investigating the 
morphology of the high [\iion{S}{ii}]/H$\alpha$ regions: photo--ionized gas will tend to be diffuse, while shocks will tend to present a 
filamentary morphology. In what follows, we assume that these two additional sources of bias are small and roughly compensate 
each other, when galaxy--integrated properties are investigated.

\subsubsection{Correlations between $L_{shock}$, $L_{shock}/L_{tot}$, $\Sigma_{SFR, HL}$, and  $M_H$}

Tables~3 and 4 report various quantities derived from the extinction--corrected \has flux and luminosity of each galaxy. These fluxes and 
luminosities are derived from the line emission images, after imposing a 5~sigma threshold to each line image, including those lines 
used for the diagnostic diagrams. While line fluxes and luminosities are corrected for the effects of dust extinction, using the  \ha/\hb line ratio, 
the diagnostic line ratios are used without extinction corrections. When E(B--V) is smaller than 0, we assign E(B--V) = 0. 

We measure two kinds of \has fluxes and luminosities: $F_{H_\alpha,tot}$ (Table~3, column~2),  which is the \has flux derived from the whole \has image, 
and $F_{H_\alpha,tot\_diag}$ (Table~3, column~4),   which is the \has flux summed over all pixels above the 5 sigma detection limit. This second 
definition mirrors the diagnostic diagram selection, which only admits pixels above the 5-sigma threshold. 
All quantities selected similarly to the diagnostic diagram pixels receive the ``diag'' subscripts.
Those ``diag'' quantities are biased measurements constrained by the detection threshold. 
We carry them along in our analysis, because the shock/photo--ionization separation itself suffers from such detection bias. 
The shock--related quantities, hence, might have better correlations with the ``diag'' quantities. 

Table~4 reports basic results related to the  shock \has luminosities, derived from the MSL, and the galaxies' SFRs. 
We calculate the SFRs from the total \has luminosities, using the relation in Kennicutt (1998); they are listed 
in the third and fourth columns in Table 4. We define the half-light area, $A_{SFR, HL}$, which is the pixel area above the half-light 
surface brightness. Using the half-light area, we define the half-light SFR density, $\Sigma_{SFR, HL}$ (column~5 of Table 4). 
We also use the diagnostic area, $A_{tot_{diag}}$, which is the total pixel 
area from the diagnostic diagram. From that diagnostic area, we calculate the SFR density, $\Sigma_{diag}$ (column~6 of Table~4).

\begin{figure}[t]
\centering
\includegraphics[height=5.0 in]{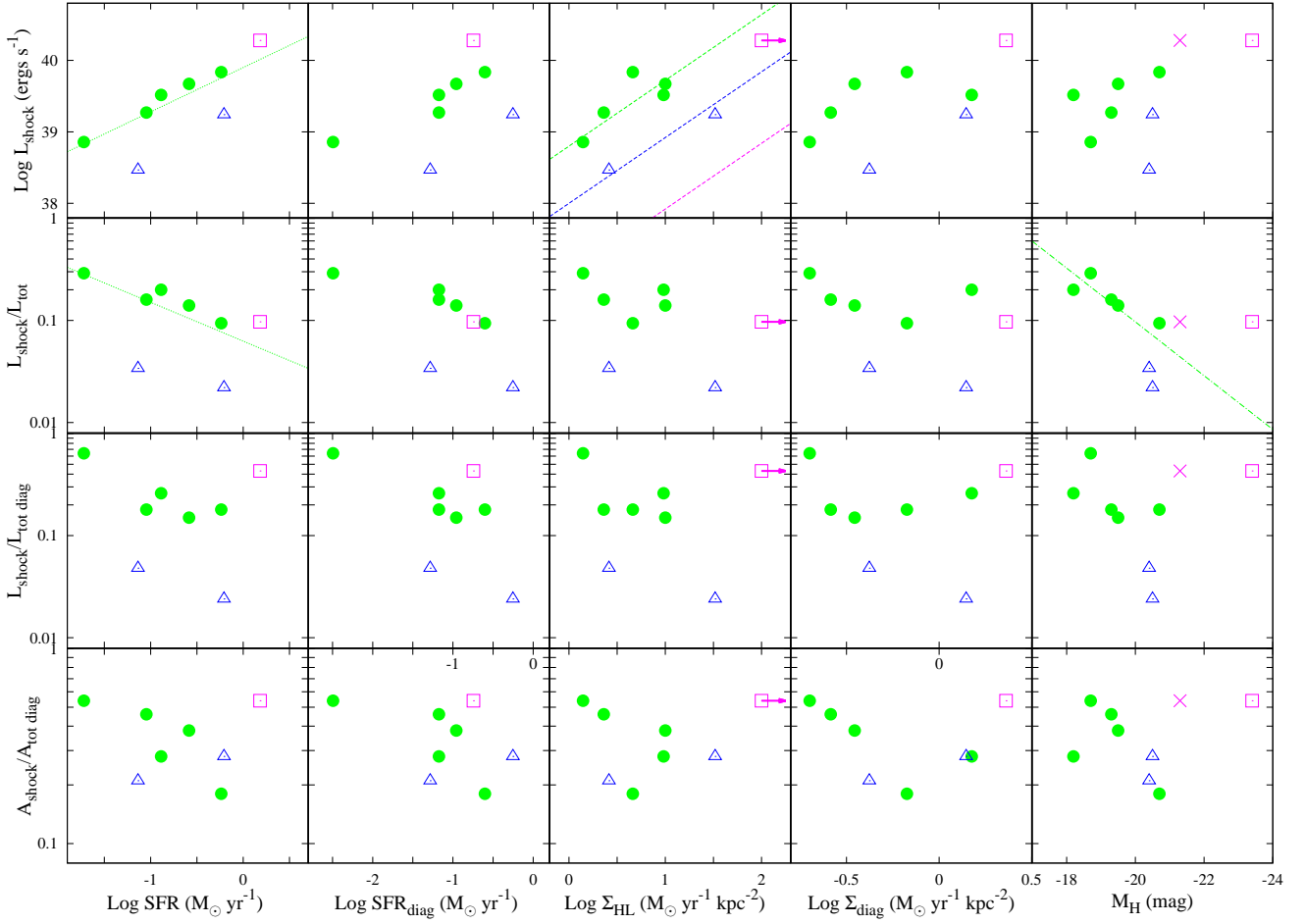}
\caption{ Relations between star formation rate (or density), absolute H-band magnitude, 
and shock--ionized gas properties presented in Table 4. 
The ``diag'' quantities, biased measurements depending on detection threshold, do not show 
any better relation than the other quantities. 
We ignore the relations connected with the ``diag'' quantities.  
The shock \has luminosity, $L_{shock}$, correlates 
with both SFR and SFR density, $\Sigma_{SFR, HL}$; the shock ratio, $L_{shock}/L_{tot}$, correlates 
with $M_H$. The green lines represent the relations presented in Equations 3 -- 6. 
The blue and magenta dashed lines on the panel of $L_{shock}$ vs. $\Sigma_{SFR, HL}$ 
show the magnitude of the offset to be applied to the green dashed line as a consquence of effects 
that lead to the underestimate of $L_{shock}$ (use of the MSL) and 
the increase of $\Sigma_{SFR, HL}$ (more compact emitting regions) in more metal rich galaxies. 
A smaller metallicity effect is observed in the $L_{shock}$ vs. SFR relation, 
because the SFR is not impacted by the size of the emitting region. 
}\label{fig:finaleleven}
\end{figure}

Figure~\ref{fig:finaleleven} summarizes the results listed in Table 4, with  
the sub--solar sample in green, the solar sample in blue, 
and the super--solar sample in magenta. 
Though many parameters seem to have no significant relation with one another, 
we find three suggestive correlations in the sub--solar group,  marked in the figure with green dotted lines. 
The green lines are chi-square minimization fits to the most refined data presented in the following subsection 
(Figure~\ref{fig:finalthirteen}) after additional corrections :
\begin{eqnarray}
\log(L_{shock}) & = & 0.62 (\pm 0.05) \times \log(SFR) \nonumber \\
& & + 39.9 (\pm 0.05) \\
& = & 0.92 (\pm 0.41) \times \log(\Sigma_{SFR, HL}) \nonumber \\
& & + 38.8 (\pm 0.26) \\
\log(L_{shock}/L_{tot}) & = & 0.26 (\pm 0.08) \times M_{H} \nonumber \\
& & + 4.25 (\pm 1.58) \\
 & = & -0.65 (\pm 0.2) \times \log (L_{H}/L_{H,\odot}) \nonumber \\
 & & + 5.11 (\pm 1.60)
\end{eqnarray}
where  $L_{tot}$ is the total \has luminosity (second column of Table 3), 
$L_{shock}$ is the total \has luminosity from shock--ionized gas component (second column of Table 4), 
$\Sigma_{SFR, HL}$ is the SFR density using the half-light pixel area (fifth column of Table 4), 
and $L_H$ is an H-band luminosity converted from $M_H$ by adopting $M_{H,\odot} = 3.32$ 
(See the next section for the details about the additional corrections and correlation tests for the equations above). 
Here we focus on the qualitative interpretation, which remains unchanged even after the application of refined corrections.
For convenience, we drop the ``\ha'' subscripts for the \has luminosities used in Table 3 and 4. 
It is interesting that all the ``diag'' quantities show worse correlations with $L_{shock}$  than the other quantities. 
This implies that the artificial detection threshold, which affects both the $L_{shock}$ and ``diag'' quantities, 
smoothes out the underlying physical relations.  Therefore, we exclude all the ``diag'' quantities hereafter, and 
$L_{shock}$ is the only quantity derived from the diagnostic diagram  
among all of the four quantities, $L_{shock}$, $L_{tot}$, $\Sigma_{SFR, HL}$, and  $M_H$ in Equation 3 -- 6. To reiterate, 
only $L_{shock}$ is a biased measurement driven by the detection threshold and by the adopted method for 
separating shocks from photo--ionized gas in diagnostic diagram.  

Since we emphasize consistency over absolute accuracy, 
we derive our two main, tantalizing qualitative results :  
\begin{enumerate}
\item $L_{shock}$ increases  with SFR and $\Sigma_{SFR, HL}$; 
\item $L_{shock}/L_{tot}$ increases as $-M_H$ decreases.   
\end{enumerate}
The first result implies that a larger energy injection from a higher SFR drives stronger radiative shocks into the surrounding ISM.
This is a quite intuitive result that we can generally expect. 
The interesting point is the sub--linear (i.e., the slope in equation~3 is lower than unity) relation between  $L_{shock}$ and SFR, implying that 
the efficiency driving radiative shocks seems not to increase as much as SFR increases. 
Another interesting point is that we find the relation between $\Sigma_{SFR, HL}$ and $L_{sh}$ too.   
The half-light area is a quantity representing the compactness of a star forming region. 
$\Sigma_{SFR, HL}$ will increase for the same SFR if the star forming region is more compact. 
This implies that, even for an identical amount of total energy injection, 
$L_{shock}$ can be larger if the injected energy is deposited in a smaller volume. 
In addition, we observe a stronger distinction between different metallicity subgroups in the $L_{shock}$ vs. $\Sigma_{SFR, HL}$ relation 
than in the  $L_{shock}$ vs. SFR relation  (top-left panel of Figure~\ref{fig:finalthirteen}). This is due to:  (1) a systematic 
underestimate of $L_{shock}$ due to the lower \oiii/\hb ratios for more metal abundant galaxies and (2) 
more compact star forming morphology for more massive galaxies (hence, higher metallicity due to mass--metallicity relation). 
This distinction is shown by the green, blue, and magenta dashed lines in Figure~\ref{fig:finalthirteen}. 

The second result shows that $L_{shock}/L_{tot}$ becomes smaller for brighter galaxies in the H--band. 
If we consider that the gravitational potential well at the center of 
galaxies is mostly shaped by the stellar mass, for which $M_{H}$ is a good approximation, 
the result shows the effect of the gravitational potential well on 
the strength of feedback-driven shocks. Higher $L_{shock}$  will be present in a shallower potential well,  
because the injected energy in a shallower potential well can be 
transported out more easily than in a deeper potential well. 

Though the quantitative values will be improved in the next section, 
the qualitative interpretations can be summarized as 
(1) $L_{shock} \propto SFR^{0.62}$ (or $L_{shock}/L_{tot} \propto SFR^{-0.38}$), 
(2)$L_{shock} \propto \Sigma_{SFR,HL}^{0.92}$, and 
(3) $L_{shock}/L_{tot} \propto 10^{0.26 M_H}$ (or $L_{shock}/L_{tot} \propto (L_H/L_{H,\odot})^{-0.65}$).

\subsubsection{The impact of detection thresholds and dust extinction}

To investigate our results in a quantitative way, 
we discuss the impact of detection thresholds and dust extinction corrections on the estimate of $L_{shock}$.  
It will turn out that this refining process for investigation yields better support for our results. 

\centerline{\it Detection thresholds}

The results shown in the previous section are based on 5 sigma detection limits for the emission line images. 
As a test, we apply two additional thresholds, 3 sigma and 7 sigma, to 
verify how much changing the detection threshold changes $L_{shock}$. In general, lowering the detection
threshold will admit more pixels in the diagnostic diagram, and increase the \has luminosity of the shock; the 
opposite happens for higher detection thresholds.

\begin{figure}[t]
\centering
\includegraphics[height=5.5 in]{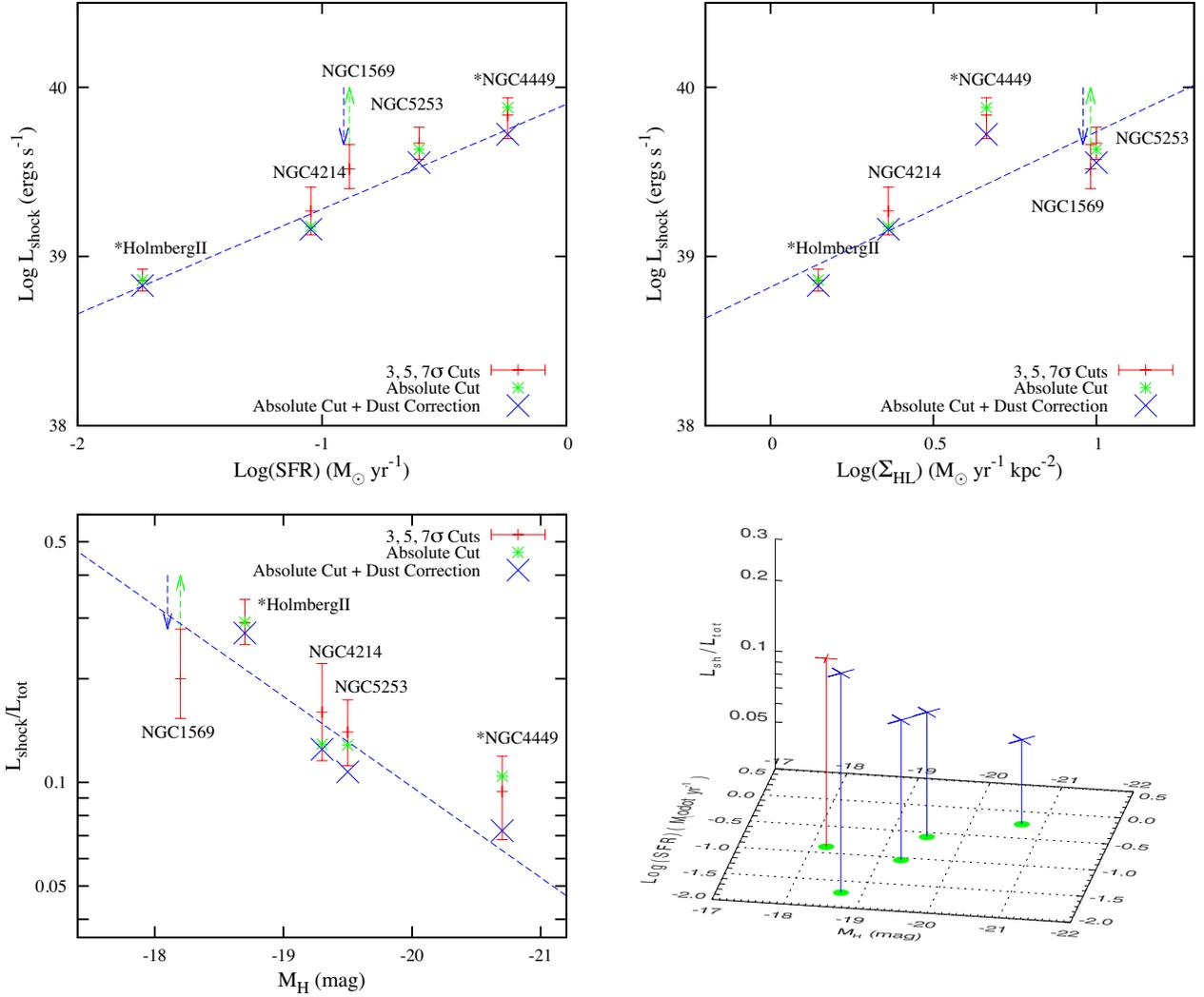}
\caption{Corrections to $L_{shock}$: apparent thresholds, $3\sigma$, $5\sigma$, and $7\sigma$ (red bar), 
absolute threshold (green point), and dust correction to the line ratios (blue point). The dashed lines represent 
$L_{shock} \propto SFR^{0.62 \pm 0.05}$, $L_{shock} \propto {\Sigma_{SFR, HL}}^{0.92 \pm 0.41}$, 
and $L_{shock}/L_{tot} \propto {10}^{0.26 (\pm0.08) M_H}$ 
(or  $ \propto {(L_H/L_{H,\odot})}^{-0.65 \pm 0.2}$). 
 Homlemberg II and NGC 4449 are marked with asterisk (*) to indicate 
their uses of the [\iion{N}{ii}] diagnostics, while the others of the [\iion{S}{ii}] diagnostics. 
From the red bars, we can find that non-negligible amount of shocks can be found from a deeper observation.  
NGC 1569 is indicated as an arrow because the absolute cut corresponds to 1 -- 2 sigma levels for the line emission images. 
Due to the heavy dust extinctions of NGC 1569, the dust correction should be significant for NGC 1569. 
We mark the dust correction effect as a blue arrow, which works in the opposite direction of the absolute cut effect. 
NGC 4449 and NGC 5253 turn out to have significant dust corrections (see the green and blue points).  
Though we have limited number statistics, this figure shows a strong evidence for the existence of regulations between the four quantities, 
$L_{shock}$, $L_{shock}/L_{tot}$, SFR, and  $M_H$.  
The bottom-right panel shows the three dimensional plot of $L_{shock}/L_{tot}$ vs. M$_H$ and SFR. 
The four blue points are the same blue points on the other panels for NGC 4449, NGC 5253, NGC 4214, and Holmberg II. 
NGC 1569 is plotted in red with its 5$\sigma$ cut result. The sub-linear scaling of $L_{shock} \propto SFR^{0.62 \pm 0.05}$ can be due to 
projection effects of the three dimensional relation. 
}\label{fig:finalthirteen}
\end{figure}

The effect of changing the threshold is shown in Figure~\ref{fig:finalthirteen}, where the $L_{shock}$ estimates for the 3, 5, 7 sigma cuts 
are reported as red points connected by vertical bars. This demonstrates  that 
the depth of the detection cut significantly affects the estimates of $L_{shock}$: $L_{shock}$ is two times larger for the 3 sigma 
threshold  than for the 7 sigma threshold. This is because the shock--ionized gas is generally  
fainter and more diffuse than the photo--ionized gas (see, C04 and H11),  and the faint pixels are those that are more 
prominently recovered by a lower threshold. Our result  suggests the existence of non-negligible amounts of shocked gas 
below our detection limit, and  emphasizes the need for an internally--consistent measurement over an accurate measurement. 

We now apply a physical threshold to the emission line images, in order to check the effect of the bias caused by the different exposure depths of 
each galaxy. We design the physical threshold by imposing a cut on the absolute surface brightness:
\begin{eqnarray}
I(\lambda)_{cut} &=& 10^{-15} ergs ~s^{-1} cm^{-2} arcsec^{-2} \\
F(\lambda)_{cut} &=& I(\lambda)_{cut} 10^{- 0.4  E(B-V)_{gas}  k_{\lambda}} \Omega_{bin},
\end{eqnarray}
with Equation~7 converting the surface brightness to a total flux. 
E(B--V)$_{gas}$ is the foreground color excess of the gas, $k_\lambda$ is the Milky Way extinction curve, 
and $\Omega_{bin}$ is the solid angle subtended by our pixel bin size. 
We set the above surface brightness threshold to the 
\ha, \hb, \oiii, and \sii images. For \nii, we use the relations, 
[\iion{N}{ii}]($\lambda\lambda6548,6583$)$~\approx~0.5\times$ \sii \msp and 
[\iion{N}{ii}]($\lambda\lambda6548,6583$)$~\approx~1.3\times$ \nii, 
for its equivalent brightness cut. We apply this threshold to all the galaxies in our sub--solar sample. 
This even physical threshold also can reduce the systematic bias from MUR spot effect presented before. 

The green points in Figure~\ref{fig:finalthirteen}, show the effect of imposing the absolute surface brightness cut. 
For NGC 4214, NGC 4449, NGC 5253, and Holmberg II, the physical threshold is equivalent to the 5 - 7 sigma detection limit. 
For NGC 1569, on the other hand, the images are relatvely shallow due to its unusually high foreground extinction,  
E(B--V) = 0.70. The physical threshold for NGC 1569 corresponds 
to 1--2 sigma detection limit, which we do not apply to the data, as far too a shallow limit. However, we report this result for NGC 1569 as an 
upward pointing green arrow in Figure~\ref{fig:finalthirteen}, to remark that our estimates are lower limits to the actual $L_{shock}$.

\centerline{\it Dust Extinction}

In general, the extinction vectors on the diagnostic diagram move the line ratios into the photo-ionized area. 
Therefore, the dust corrected estimates of $L_{shock}$ decrease after dust extinction correction. 
Since the extinction vector is small for the diagnostic diagram, the dust extinction corrections 
generally produce minor effects on the line ratios. 
But if many data points are crowded near our shock separating line, 
even the small extinction vector can migrate non-negligible amount pixels from shock--ionized into photo-ionized area. 
We find that the two  galaxies, NGC 4449 and NGC 1569, have relatively high dust corrections 
and NGC 5253 has a non-negligible dust correction.  

\begin{figure}[t]
\centering
\includegraphics[height=2.5 in]{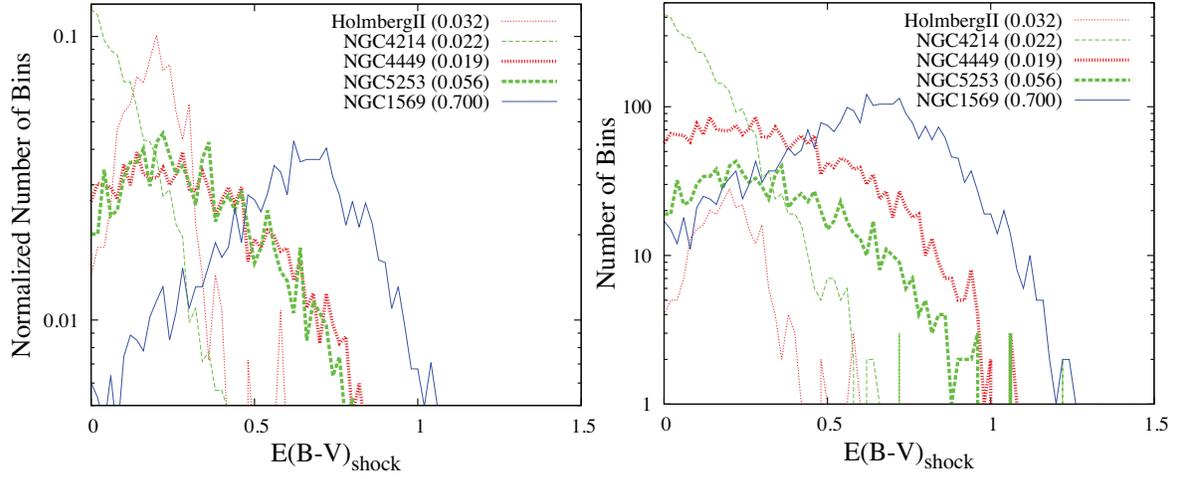}
\caption{The dust extinction, E(B--V), applied to shock--ionized bins. 
The right panel shows the absolute counts of the shock--ionized bins to which  dust correction is applied 
and the left panel the normalized counts. 
Holmberg II and NGC 4214 show small dust extinction, while NGC 1569 shows very high E(B--V) due to 
its heavy foreground extinction. 
The dust extinction correction are not negligible in NGC 5253, NGC 4449 and NGC 1569. 
NGC 5253 and NGC 4449 show similar distributions of E(B--V) in the left panel. But, in the right panel, 
we find that NGC 4449 has more pixel bins than NGC 5253. Due to this, as described in \S 4.2.2, 
the dust correction affects proportionally more NGC 4449 than NGC 5253. 
}\label{fig:finalfourteen}
\end{figure}

Figure~\ref{fig:finalfourteen} shows the distribution of E(B--V) color 
excesses used for each galaxy on a pixel--by--pixel basis. 
The right panel shows the absolute counts of the shockÐionized bins for dust correction 
and the left panel shows the normalized counts. The foreground extinction values are given with the names in the round braces. 
As mentioned previously, NGC 1569 suffers from heavy foreground dust extinction, hence it shows the most significant color excess.  
On the other hand, NGC 4449 and NGC 5253 have relatively high internal dust extinctions, 
while their foreground dust extinctions are much smaller than NGC 1569; E(B--V) = 0.019 for NGC 4449 and 0.056 for NGC 5253. 
The other two galaxies, Holmberg II and NGC 4214, show minor dust extinctions. 
We can expect that the dust correction will affect the shock estimates for NGC 4449, NGC 5253, and NGC 1569.

The blue points in Figure~\ref{fig:finalthirteen} show the estimates of shocks 
after dust corrections are applied to the diagnostic diagrams. 
We can readily observe that the dust corrections on the diagnostic diagrams have some effect on 
NGC 4449 and NGC 5253. For NGC 1569, we mark the dust correction effect as a blue arrow, since the dust correction will 
work in the opposite direction (reducing the shock estimate) of the physical threshold. And the correction should be the most for NGC 1569.  
For NGC 4214 and Holmberg II, the dust extinctions are small. Both galaxies are relatively free from all of the issues 
related to dust extinction correction. 
NGC 4449 and NGC 5253 show a similar color excess on the left panel (for the normalized count of bins) in Figure~\ref{fig:finalfourteen}. 
This shows that they have a similar dust extinction property.  However, NGC 4449 is brighter than NGC 5253 in \has emission; 
hence, the dust correction affects more the estimates of bins containing shocked gas in  NGC 4449 than in NGC 5253. 
When considering the error propagation by dust correction, 
the reliability of correction for NGC 4449 and NGC 5253 could be arguable. 
At least, however, we can find that the dust correction does not change our qualitative results, and it seems to make our correlations even tighter. 

To summarize, the blue points are our final shock luminosity estimates, with the caveat that dust extinction corrections 
are in general uncertain.  From fitting the blue points, we obtain Equations~3 -- 6 , whose statistical reliability is given 
via correlation tests in Table~5. 
All the effects presented so far  induce vertical offsets in the relations of equations~3 -- 6,  but 
the slopes will be generally minimally affected, and remain robust against these effects.

\section{Discussion}

\subsection{Interpretation of $L_{shock}/L_{tot} $}

The deposited energy from stellar feedback, $E_{mech}$, cools away by various cooling mechanisms, $E_{loss}$. 
The remaining energy, $E_{wind}$, can drive galactic scale winds : 
\begin{equation}
E_{wind}(t) = E_{mech} (t) - E_{loss}(t)
\end{equation}
The amount of feedback energy deposited into the surrounding ISM (and its rate) can be calculated from stellar population synthesis models (STARBURST99 in this paper) 
when we have (or assume) a star formation history, $h_\star(t)$, 
\begin{eqnarray}
E_{mech}(t) &=& \int_{-\infty}^{t} L_{mech} (t^\prime) dt^\prime \\
L_{mech}(t) &=& \int_{-\infty}^{t} K_{inst} (t-t^\prime) h_\star (t^\prime) dt^\prime,
\end{eqnarray}
where $K_{inst}(t)$ is a kernel function of luminosity evolution for instantaneous star formation. Many models and assumptions 
are involved in calculating the luminosity such as the metallicity, initial mass function (IMF), stellar atmosphere models, and stellar evolution tracks. 
We need to keep in mind this complexity when modeling and interpreting stellar feedback. 

The two models most  commonly used are the 
instantaneous burst, $h_\star (t) = \tilde{h}_\star \delta(t - 0)$, and the continuous star formation, $h_\star (t) = \tilde{h}_\star \theta(t - 0)$, 
here written mathematically using the conventional delta function and step function. The mechanical luminosity from stellar feedback for each 
star forming model can be written as : 
\begin{eqnarray}
L_{inst}(t) & = & \tilde{h}_\star K_{inst}(t) \\
L_{cont}(t) & = & \tilde{h}_\star \int_{0}^{t} K_{inst} (t-t^\prime) dt^\prime \\
 & \approx & \tilde{h}_\star K_0 ~~~~~~~ (t > 40 Myr)
\end{eqnarray}
The two luminosity templates, $K_{inst}(t)$ and $K_{cont}(t) \equiv \int_{0}^{t} K_{inst} (t-t^\prime) dt^\prime$, 
can be provided by STARBURST99 (Leitherer et al. 1999). One of important results from the templates is 
that, after 40 Myr, the luminosity of the continuous star formation model is stabilized to a constant, $K_0 \approx 10^{42} (ergs ~ s^{-1})$, 
while the luminosity of the instantaneous model fades out to $ < 10^{36} (ergs ~s^{-1})$, 
because all massive stars explode as supernovae within 40 Myr. 

Because the hydrogen recombination lines are tracers of star formation rate 
in a very short term period ($< 10$ Myr; Leitherer et al. 1999), the obtained SFR can be considered as an instantaneous measure 
of current star formation. Ideally, we have to subtract $L_{shock}$ from $L_{tot}$ to obtain $h_\star$ 
(as a reminder, $L_{shock}$ and $L_{tot}$ are \has luminosities).  
Because $L_{shock}$ is a small fraction of $L_{tot}$ 
for most normal star forming galaxies and it is generally hard to measure $L_{shock}$, as explained through this paper, 
conventionally $L_{shock}$ has been ignored : 
\begin{eqnarray}
h_\star &=& \kappa (L_{tot} - L_{shock}) \\
  &\approx & \kappa L_{tot}  ~~~~~~~~~~~~; \kappa = 7.9 \times 10^{-42}
\end{eqnarray}
In our sample,  the fraction of $L_{shock}$ is at most 0.3, so the conventional approximation would be  acceptable also in our 
case. We keep, however, the explicit formula, Equation 15, which will be used later to derive $L_{shock}/L_{tot}$. 
From the observed SFR, $h_\star$, we can rewrite the mechanical luminosity of continuous burst model after 40 Myr, 
\begin{equation}
L_{mech} = \kappa (L_{tot} - L_{shock}) K_0 \approx \kappa L_{tot} K_0  ~~~ (t > 40 Myr)
\end{equation}
In general, $K_0$ can be used in place of the time-dependent templates, $K_{inst}(t)$ or $K_{cont}(t)$. 

Gas cooling is more complicated to estimate than stellar feedback energy because 
thermodynamic and hydrodynamic interactions are involved between surrounding gas and feedback energy.   
In the early stages of star formation, the energy deposited into the ISM produces a hot bubble that  
expands adiabatically (Weaver et al. 1977). Hence, most of the feedback energy is stored 
as $E_{wind}$ and is invisible in the optical. As the hot bubble adiabatically cools, 
the expanding shock front becomes radiative. Our measured $L_{shock}$ is a tracer of 
this radiative shock. The total radiative loss, $L_{shock loss}$, can be estimated from the observed \has loss, $L_{shock}$,   
\begin{equation}
L_{shockloss} = \lambda L_{shock}
\end{equation}
The conversion factor, $\lambda$, depends on the shock models and ISM properties, such as metallicity, ambient gas density, 
magnetic field strength, and pre-ionization fraction. Rich et al. (2010) estimate $\lambda = 20 - 80$ for slow radiative shocks. 
We use this range of values for our qualitative interpretation, but note  the complexity of treating radiative shocks, which includes the  
uncertainties in the parameters chosen for the population synthesis models. 

\begin{figure}[t]
\centering
\includegraphics[height=6.0 in, angle=270]{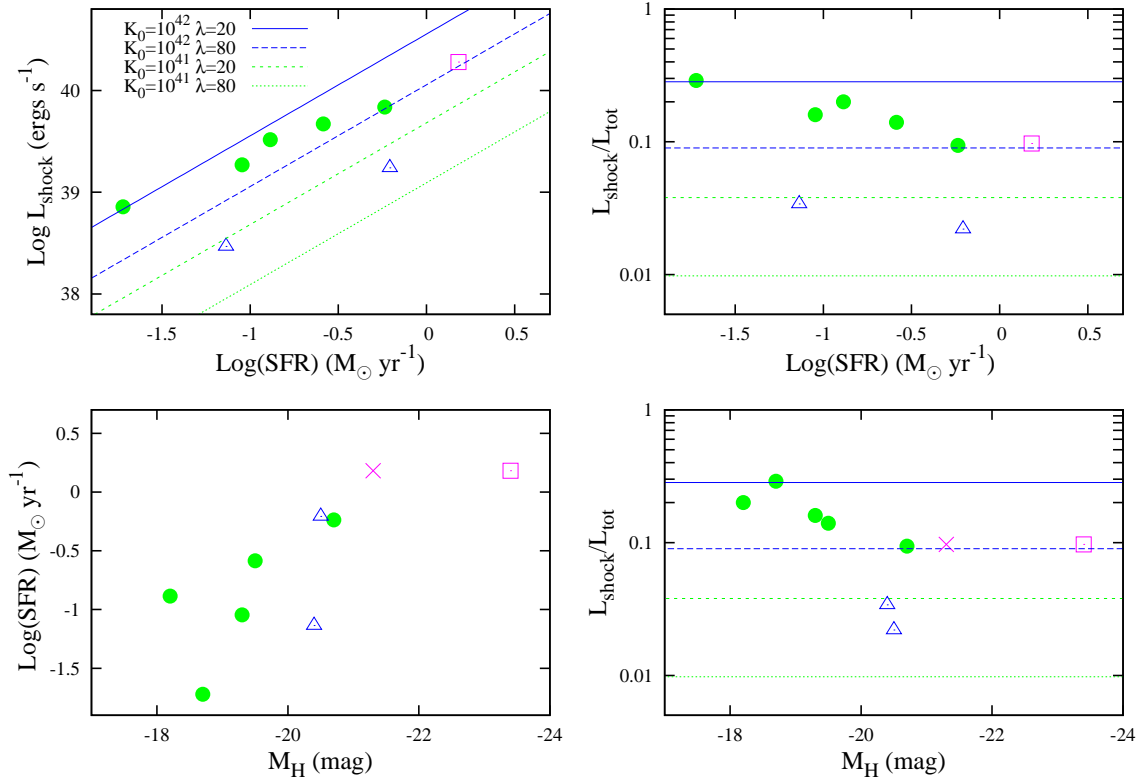}
\caption{ Relations between SFR -- $L_{sh}$(top-left), SFR -- $L_{sh}/L_{tot}$(top-right), 
$M_H$ -- SFR(bottom-right), and $M_H$ -- $L_{sh}/L_{tot}$(bottom-right), 
presented in Figure~\ref{fig:finalone} and Figure~\ref{fig:finaleleven}. 
Due to SFR$= \kappa L_{tot}$, 
the two top panels are equivalent. The lines represent the theoretical ratio $\mu/(1+\mu)$, 
where $\mu = \kappa K_0/\lambda$ in Equation 19 and 20. 
The range of $\mu$ can cover the observed data. 
Because $M_H$ -- SFR is not independent (Noeske et al. 2007), 
the relations on the right panels are projected results from the underlying 3 dimensional relation 
of $L_{sh}/L_{tot} \sim (M_H, $SFR). The bottom-right panel of Figure~\ref{fig:finalthirteen} shows the preliminary 3 dimensional relation. 
}\label{fig:finaltwelve}
\end{figure}

Now we derive an estimate of $L_{shock}/L_{tot}$ from Equation 17 and 18 : 
\begin{eqnarray}
\frac{L_{shock}}{L_{tot}} &=& \frac{\mu}{1+\mu} \approx \mu \\
\mu &\equiv& \big( \frac{\kappa K_0}{\lambda} \big)\big( \frac{L_{shockloss}}{L_{mech}} \big)
\end{eqnarray}
The term, $ \frac{\mu}{1+\mu}$, is the explicit derivation from $h_\star = \kappa (L_{tot} - L_{shock})$ and 
can be approximated to $\mu$ when $\mu$ is small enough, $\mu < 0.1$.  Our observed ratios, $L_{shock}/L_{tot} = 0.05 - 0.40$, justify 
the approximation, as we discuss below. 
When we take the fiducial values, $K_0 = 10^{42}$ and $\lambda = 60$, we obtain $ \big( \frac{\kappa K_0}{\lambda} \big) = 0.13$. 
If we allow a larger range of  values to cover most physical conditions, $K(t) = 10^{36 - 42}$ and $\lambda = 20 -80$, 
we obtain $ \big( \frac{\kappa K_0}{\lambda} \big) = 0.00 - 0.40$, which is in agreement with the observed range of ratios for  
$L_{shock}/L_{tot} = 0.05 - 0.40$. Indeed, as shown in Figure~\ref{fig:finaltwelve}, observed range and  
theoretical expectations  overlap, implying that  in our sample $L_{shockloss}/L_{mech}\approx 1$. 
 
In order to attempt an explanation of the scaling relation $L_{shock}/L_{tot} \propto {(L_H/L_{\odot,H})}^{-0.65}$, we discuss 
the three parameters, $K_0$, $\lambda$, and $L_{shockloss}/L_{mech}$, in greater detail. $\kappa$ is dictated 
by atomic physics and will not be discussed further. 
As presented above, $K_0$ depends on the IMF, stellar atmosphere models, stellar evolution models, stellar metallicity, and star formation history, 
while $\lambda$ is a function of the shock model and the ISM properties. As we have divided our sample in sub--samples according 
to metal content, we can assume that variations in stellar metallicity and ISM properties are minimized. Furthermore, IMF, stellar 
atmosphere models and evolution models are not (likely) a function of the galactic environment. The star formation history of our galaxies 
has remained constant over the past few tens of Myr at least, thus $K_0$ is likely to have remained roughly constant. If we assume that 
the dependency of $\lambda$ over the specific shock model is small, then the role of $ \big( \frac{\kappa K_0}{\lambda} \big)$ in Equation~20 
is to simply set the absolute scale of the relation. Much of the environment dependency, i.e., the dependency on L$_H$, is carried by 
$L_{shockloss}/L_{mech}$. 

The ratio $L_{shockloss}/L_{mech}$ represents the energy balance between the gain L$_{mech}$ from feedback energy and 
loss $L_{shockloss}$ by radiative shock. The higher the value the higher the loss of feedback energy through radiation. 
To drive a gas outflow which may develop into a galactic super wind, the hot bubble should retain sufficient kinetic energy to expand 
to a scale height comparable to or larger than that of the galactic disk (e.g. MacLow et al. 1989, de Young \& Heckman 1994, 
Murray et al. 2010). For our galaxies, we have a high likelihood that no such major gas outflow will be driven out of the galaxy: we derive 
$L_{shockloss}/L_{mech}\approx 1$, although it  only represents the current strength of underlying gas expansion driving radiative shocks. 
The case of $L_{shockloss}/L_{mech} \ge 1$ is very unlikely, when we include other channels of energy loss, such as X-ray emissions. 
But still it is not impossible since a lot of energy can be cumulated during early adiabatic phase, 
then released through radiative shocks in a short period of time. Overall, there is a strong argument for the environment (stellar mass) dependency in 
our scaling relation to be carried by $L_{shockloss}/L_{mech}$.

\subsection{$L_{shock}/L_{tot} \propto  (L_{H}/L_{H,\odot})^{-\alpha} (SFR)^{-\beta} $}

Given our limited sample size and the radiative nature of the stellar feedback we observe, the content of this section 
is speculative at best, but provides some tantalizing suggestions and a direction for improvement and  progress. 

Our two scaling relations:  (1) $L_{shock} \propto {SFR}^{~0.62}$ and (2) $L_{shock}/L_{tot} \propto {(L_H/L_{\odot,H})}^{-0.65}$, are 
suggestively similar to those derived in a recent simulation by  Hopkins et al. (2012). These authors derive relations between the  wind mass-loss rate driven by stellar feedback, $\dot{M}_{wind}$, and the two quantities of star formation rate, $\dot{M}_{\star}$, and stellar mass, ${M}_{\star}$: 
(1) $\dot{M}_{wind} \propto{\dot{M}_{\star}}^{0.7}$ and (2) $\dot{M}_{wind}/\dot{M}_{\star} \propto (M_{\star}/M_{\odot})^{- (0.25 - 0.5)}$. 
The scaling relations derived by Hopkins et al. are for feedback-driven galactic winds, while our relations are for radiative shocks, so the 
similarities should be taken with caution at this stage. However, if the way in which stellar feedback scales with both SFR and stellar mass 
is independent of the fate of the feedback energy, we can postulate that our observed scaling relations, derived for radiative shocks, reproduce 
those derived from simulations of end-phase galactic super winds. 

Conversely, also the opposite argument could be made. Our observational scaling relations indicate 
that radiative losses, L$_{shock}$/L$_{tot}$, decrease for increasing stellar mass. In more massive galaxies the level of radiative losses is small so 
that most of the stellar feedback energy then is available to drive gas motions, and, possibly, a wind.  On the other hand, we find that radiative losses in dwarf galaxies are substantial and, thus, may reduce the efficiency of driving mass loss. The effects of radiative losses, therefore, go in an opposite sense to 
the depth of the gravitational potential wells: more massive galaxies have deeper potentials, but also significantly smaller fractional radiative losses 
of mechanical energy. Which of the two interpretations, the similarity with the Hopkins et al.'s result or the opposite trend,  is likely to be correct 
is unclear at this point.

Another important issue is related to the fact that SFR and stellar mass are not independent (Noeske et al. 2007, see also Figure~\ref{fig:finaltwelve} 
above). Thus, the measured scaling exponents, 
$L_{shock}/L_{tot} \propto {SFR}^{-0.38}$ (or $L_{shock} \propto {SFR}^{~0.62}$) and $L_{shock}/L_{tot} \propto {(L_H/L_{\odot,H})}^{-0.65}$, 
are the projected values from the real exponents.  
The actual scaling relation is likely to be expressed as a combination of factors: 
\begin{equation}
L_{shock}/L_{tot} \propto  (L_{H}/L_{H,\odot})^{-\alpha} (SFR)^{-\beta} 
\end{equation}
Our limited sample size currently prevents us from performing a reliable multi-dimensional fit for deriving $\alpha$ and $\beta$.
However, a tentative planar fit using the 4 points in the bottom-right panel of Figure~\ref{fig:finalthirteen}, 
yields the values $\alpha \approx 0.67$ and $\beta \approx 0.01$. Although these are preliminary results, we can speculate that the 
observed sublinear relation:  $L_{shock} \propto {SFR}^{~0.62}$, is the result of a projection effect, as the 
actual relation:  $L_{shock} \propto SFR^{(1 - \beta)}$ appears to have an exponent close to unity. 
Another important consideration is  the robustness of the relation between $L_{shock}/L_{tot}$ and stellar mass:  
the preliminary exponent $\alpha$ from the planar fit is 0.67, while the projected exponent is 0.62. If a more extensive study involving 
a larger sample of galaxies confirms $\beta\sim$0, we can conclude that the stellar mass scaling is the dominant modulator 
for $L_{shock}/L_{tot}$. We should note that Hopkins et al. (2012) combine the scaling relations in their simulations into a single multi--parameter 
equation, and they also recover a linear exponent, $\dot{M}_{wind} \propto{\dot{M}_{\star}}$, between the wind mass-loss rate and the SFR.

\section{Summary}

With HST narrow-- and broad--band imaging, 
we have identified stellar--feedback induced shocks in a sample of 8 local starburst galaxies 
and investigated the relations among shock-related quantities, stellar mass, and 
star formation. We perform our analysis in a spatially resolved fashion, sampling our galaxies 
in pixels about 3--15~pc in size. Though our number statistics are still limited, we have found some 
indication of a correlation between the observed shock--ionized gas and 
the star forming activity. In summary:    

\begin{enumerate}
\item After dividing our sample into three subgroups according to their metallicities: 
sub--solar  (NGC 1569, NGC 5253, NGC 4449, Holmberg II, NGC 4214), 
solar  (He 2-10, NGC 3077), and super--solar (NGC 5236), we apply the 
 ``maximum starburst line" in an internally--consistent manner to separate shock--ionized gas from 
 photo--ionized gas in a classical diagnostic plot of the [\iion{O}{iii}]/H$\beta$ ratio versus
  the [\iion{S}{ii}]/H$\alpha$ (or [\iion{N}{ii}]/H$\alpha$) ratio.  The fraction of pixels identified as shock--dominated 
 systematically decreases for increasing metallicity, indicating that 
 shocks are underestimated  in higher metallicity galaxies when the maximum starburst line is applied.  
 A common problem to our analysis is that weak shocks projected on strongly photo--ionized regions will 
 not be identified, likely leading to some underestimate of the total shock luminosity in a galaxy. 
\item The [$\iion{S}{ii}$] diagnostic is preferred for shock discrimination to the [$\iion{N}{ii}$] diagnostic, 
because it has less overlap between the photo-ionized and the shock--ionized components. 
Conversely, the [$\iion{N}{ii}$] diagnostics is more sensitive to metallicity than the [$\iion{S}{ii}$] diagnostics.  
\item The distribution of single line ratios is consistent with theoretical expectations. 
Due to its metallicity dependence, the [$\iion{O}{iii}$]/\hb ratio provides poor discrimination  
between shocks and photo--ionized gas, especially for metal poor galaxies. 
Both [$\iion{N}{ii}$]/\has and [$\iion{S}{ii}$]/\has can be used as single--line ratios for estimating shocks, with  
the [$\iion{S}{ii}$]/\has ratio providing a better shock discriminator, because of its lower degree of degeneracy for varying 
physical conditions. In general, two line ratio diagnostics should be preferred over single line ratio 
diagnostics, because the latter are sensitive not only to shocks, but also to hot, photo--ionized gas, thus leading to 
an overestimate of the shock luminosity. However, morphology can help discriminate between the two cases, since 
photo--ionized gas is more uniformly diffused, while  shock--ionized gas has a more filamentary structure.
\item  We find that: (1) a larger \has luminosity from shocks, L$_{shock}$, is found for higher star formation rate 
with a sub--linear scaling; and (2) a higher ratio of the shock to the total \has luminosity,  $L_{shock}/L_{tot}$, 
is obtained for increasing absolute H-band magnitudes,  $M_H$ (lower 
stellar masses). The two scaling relations are expressed as: 
$L_{shock} \propto {SFR}^{~0.62}$ and $L_{shock}/L_{tot} \propto ( {L_H/L_{\odot,H} )}^{-0.65}$. 
These results have been obtained 
for the sub--solar sample, NGC 1569, NGC 5253, NGC 4449, Holmberg II, and NGC 4214. No similar conclusions can be derived at 
this point for the other metallicity groups, due to the small sample statistics, although their trends are consistent with 
those of the sub--solar sample. We find tantalizing similarities between our observationally derived scaling relations and those 
recently obtained from simulations of galactic super winds. 
\item Our results are derived for starburst galaxies, i.e., for galaxies where we expect the effects of stellar feedback to have 
the most impact on the surrounding environment. Quiescently star--forming galaxies can be expected to have less 
substantial supernova powered winds. 
\end{enumerate}

Due to the limited sample statistics, our results need to be refined by larger samples, especially in the high metallicity bins.  
However, we find evidence for the existence, in starburst galaxies, of regulation of the strength of radiative shocks 
by two galactic parameters: the star formation rate and the stellar mass. 

\acknowledgments
S.H. and D.C. acknowledge support from grants associated with programs \# 10522, 11146, and 11360,  
provided by NASA through  the Space Telescope Science Institute, which is operated by 
the Association of Universities for Research in Astronomy, Inc., under NASA contract NAS 5-26555. 
C.L.M acknowledges support from NSF under grant AST-1109288. 

This paper is based on observations made with the NASA/ESA Hubble Space Telescope, 
obtained  at the Space Telescope Science Institute, which is operated 
by the Association of Universities for Research in Astronomy, Inc., under NASA contract NAS 5-26555. 
These observations are associated with program \# 10522 and 11146. 
The paper also uses Early Release Science observations made by the WFC3 Science Oversight Committee (program \# 11360). 
We are grateful to the Director of STScI for awarding DirectorÕs Discretionary time for this program.

\clearpage

\begin{deluxetable}{lcrrrrr}
\tabletypesize{\footnotesize}
\tablecolumns{7}
\tablewidth{0pc}
\tablecaption{Basic Information about the Sample Galaxies}
\tablehead{
\colhead{Galaxy\tablenotemark{$\dagger$}}& \colhead{Instrument\tablenotemark{a}}& \colhead{Bin size\tablenotemark{b}} 
&\colhead{[\iion{N}{ii}] corrections\tablenotemark{c}} 
& \colhead{$12+\log$(O/H)\tablenotemark{d}} & 
\colhead{D\tablenotemark{e}} & \colhead{M$_H$\tablenotemark{f}} \\
\colhead{} &\colhead{} &\colhead{physical(angular)} & \colhead{} & 
\colhead{} &\colhead{(Mpc)} & \colhead{(mag)}  
}
\startdata
NGC 1569  & ACS+WFPC2 & 2.8pc ($0.3''$)& [\iion{N}{ii}]/\ha = 0.15 & 8.19  & 1.90(E) & -18.2\\
NGC 5253   & WFPC2 &  5.8pc ($0.3''$)& [\iion{N}{ii}]/[\iion{S}{ii}] $\approx 0.5$  &8.19  & 4.00(C) & -19.5\\
NGC 4214   & WFC3 &  2.9pc ($0.2''$)& [\iion{N}{ii}]/[\iion{S}{ii}] $\approx 0.5$  &8.20  & 2.94(C) & -19.3\\
He 2-10   & WFPC2 &  13pc ($0.3''$)& [\iion{N}{ii}]/\ha = 0.18  &8.40  & 9.0(E) & -20.5 \\
NGC 3077   & WFPC2 & 5.6pc ($0.3''$) & [\iion{N}{ii}]/[\iion{S}{ii}] $\approx 1.0$  &8.64  & 3.82(K) & -20.4 \\
NGC 5236(M83)  & WFC3 & 4.3pc ($0.2''$)& [\iion{N}{ii}]/\ha = 0.54 &8.94 & 4.47(K) & -23.4 \\
\cline{1-7}\\
Holmberg II   & ACS+WFPC2 & 4.1pc ($0.25''$)& ...     &7.92 & 3.39(M) & -18.7 \\
NGC 4449   & ACS+WFPC2 & 5.1pc  ($0.25''$)& ...   &8.22 &  4.21(K)& -20.7 \\
\enddata
\tablenotetext{a}{Instrument used for narrow(or broad) -band observations. 
Because our sample is a collection of several GO programs, various instruments have been used.} 
\tablenotetext{b}{Physical and angular (in parentheses) bin size used in the diagnostic diagrams. } 
\tablenotetext{c}{ Adopted method for [\iion{N}{ii}] correction. The [\iion{N}{ii}]/\ha ~ratios are adopted from Kennicutt et al. (2008) and Moustakas et al. (2010). 
The estimated [\iion{N}{ii}]/[\iion{S}{ii}] ratios from Martin (1997) are 0.3--0.5 for NGC 5253, 0.7 for NGC 4214, and 1.0 for NGC 3077. 
Hence, we choose the fractions, $\approx 0.5$ for the sub-solar group and $\approx 1.0$ for the solar group. 
C04 indicate that  the differences are only 2--4\% between the two corrections for NGC 5253, NGC 4214, and NGC 3077, 
while the difference is significant for NGC 5236 (M83). At least, therefore, the [\iion{N}{ii}] correction within each subgroup is consistent. } 
\tablenotetext{d}{ Oxygen abundances from Kobulnicky \& Skillman (1996; NGC 4214 and  NGC1569), Kobulnicky et al. (1997; NGC 5253), 
Kobulnicky et al. (1999; NGC 5253, NGC 4214, and He 2-10), Pilyugin et al. (2004; NGC 4214, Holmberg II, and NGC 5236), 
Bresolin et al. (2005; the center of NGC 5236), Croxall et al. (2009; Holmberg II and NGC 3077), Bresolin (2011;  NGC 5253 and NGC5236), 
Gueseva et al. (2011; NGC 5253 and He 2-10), and Berg et al. (2012; NGC 4449). Multiple measurements are averaged. 
For NGC 5236, the metallicity in the central region is presented.  
The metallicity of He 2-10 is slightly subsolar, but its diagnostic diagram is more similar with NGC3077 than the sub-solar galaxies. 
This implies that the measured shock is more biased when compared with the sub-solar group than with the solar group.
} \tablenotetext{e}{ Distances from C04 ©, Kennicutt et al. 2008 (K), Engelbracht et al. 2008 (E) and Moustakas et al. 2010 (M)} 
\tablenotetext{f}{Absolute H-band magnitude, from 2MASS.}
\tablenotetext{$\dagger$}{For six galaxies, NGC 1569, NGC 5253, NGC 4214, He 2-10, NGC 3077, and NGC 5236 
the [\iion{S}{ii}] emission line was observed; for Holmberg II and NGC 4449, we obtained the [\iion{N}{ii}] emission line.}
\end{deluxetable}

\begin{deluxetable}{lllclrrr}
\tabletypesize{\scriptsize}
\tablecolumns{8}
\tablewidth{0pc}
\tablecaption{Summary of the narrow band observations.}
\tablehead{
\colhead{Galaxy\tablenotemark{$\dagger$}} 
& \colhead{z} 
& \colhead{Filter} 
& \colhead{Line}
& \colhead{Continuum} 
& \colhead{Program ID}  & \colhead{EXPTIME} & \colhead{1$\sigma$ limit}  \\
\colhead{}  & \colhead{}  & \colhead{}   & \colhead{}  & \colhead{}  & \colhead{}  
& \colhead{(sec)}  & \colhead{($ergs ~ s^{-1} cm^{-2}$)}  
}
\startdata
He 2-10 & 0.002912 & F658N(WFPC2) & H$\alpha$ + [N {\tiny{II}}] & F547M F814W & 11146 & 2000 &$6.3 \times 10^{-18}$\\
 & & F487N(WFPC2) & H$\beta$ & F547M  & 11146  & 4400 &$1.0 \times 10^{-17}$\\
 & & F502N(WFPC2) & [O \tiny{III}] & F547M & 11146  & 5200& $8.0 \times 10^{-18}$\\
 & & F673N(WFPC2) & [S \tiny{II}] & F547M F814W & 11146 & 3200&$4.7 \times 10^{-18}$ \\
 \cline{1-8}\\
NGC 1569 & -0.000347 & F656N(WFPC2) & H$\alpha$ + [N {\tiny{II}}] & F547M F814W  & 8133*  & 1600 &$1.2 \times 10^{-17}$\\ 
 & & F487N(WFPC2) & H$\beta$&  F547M  & 8133*  & 2400 &$1.7 \times 10^{-17}$\\
 & & F502N(WFPC2) & [O \tiny{III}] & F547M  & 8133*  & 1500 &$2.1 \times 10^{-17}$\\
 & & F673N(WFPC2) & [S \tiny{II}] & F547M F814W & 8133*  & 3000 &$ 1.3\times 10^{-17}$\\
\cline{1-8}\\
NGC 4449 & 0.000690 & F658N(ACS) & H$\alpha$ + [N {\tiny{II}}] & F550M F814W  & 10585*  & 1539 &$3.8 \times 10^{-18}$\\
 & & F660N(ACS) & [N {\tiny{II}}] + H$\alpha$ &  F550M F814W & 10522  & 1860 &$1.3 \times 10^{-17}$\\
 & & F487N(WFPC2) & H$\beta$& F550M  & 10522 & 2100 &$8.0 \times 10^{-18}$\\
 & & F502N(ACS) & [O \tiny{III}] &  F550M  & 10522 & 1284 &$1.0 \times 10^{-17}$\\
\cline{1-8}\\
Ho-2 & 0.000474 & F658N(ACS) & H$\alpha$ + [N {\tiny{II}}] & F550M F814W & 10522 & 1680 & $1.0 \times 10^{-17}$\\
 & & F660N(ACS) & [N {\tiny{II}}] + H$\alpha$ & F550M F814W  & 10522 & 1686 &$6.4 \times 10^{-18}$\\
 & & F487N(WFPC2) & H$\beta$ & F550M  & 10522 & 5400 & $6.6 \times 10^{-18}$\\
 & & F502N(ACS) & [O \tiny{III}] & F550M  & 10522 & 1650 &$7.2 \times 10^{-18}$\\
\cline{1-8}\\
NGC 5236 & 0.001711 & F657N(WFC3) & H$\alpha$ + [N {\tiny{II}}] &  F555W F814W & 11360 & 1484 & $6.1 \times 10^{-18}$ \\
 & & F487N(WFC3) & H$\beta$ & F555W & 11360 &  2700 &$4.3 \times 10^{-18}$\\
 & & F502N(WFC3) & [O \tiny{III}] & F555W  &11360 &  2484 &$4.6 \times 10^{-18}$\\
 & & F673N(WFC3) & [S \tiny{II}] &  F555W F814W & 11360 & 1850 &$5.8 \times 10^{-18}$\\
\cline{1-8}\\
NGC 4214 & 0.000970 & F657N(WFC3) & H$\alpha$ + [N {\tiny{II}}] & F555W F814W &11360 & 1592 & $5.3 \times 10^{-18}$\\
 & & F487N(WFC3) & H$\beta$ & F555W & 11360  & 1760 &$5.8 \times 10^{-18}$\\
 & & F502N(WFC3) & [O \tiny{III}] & F555W &11360  & 1470 &$6.2 \times 10^{-18}$\\
 & & F673N(WFC3) & [S \tiny{II}] &  F555W F814W & 11360 & 2940 &$3.5 \times 10^{-18}$\\
\enddata
\tablenotetext{$\dagger$}{NGC 3077 and NGC 5253 are absent here because we adopt 
the results for the galaxies from Calzetti et al. (2004; GO--9144). 
The starred program IDs are open access not related to our programs. }
\end{deluxetable}

\begin{deluxetable}{lrrrr}
\tabletypesize{\tiny}
\tablecolumns{5}
\tablewidth{0pc}
\tablecaption{Measured and Derived H$\alpha$ Quantities}
\tablehead{
\colhead{Galaxy} & \colhead{$F_{H_\alpha,tot}$ \tablenotemark{a}} 
& \colhead{$L_{H_\alpha,tot}$\tablenotemark{b}} 
& \colhead{$F_{H_\alpha,tot\_diag}$ \tablenotemark{c}} 
& \colhead{$L_{H_\alpha,tot\_diag}$\tablenotemark{d}}\\
\colhead{}  & \colhead{ (erg s$^{-1}$ cm$^{-2}$)} &\colhead{(erg s$^{-1}$)} 
& \colhead{ (erg s$^{-1}$ cm$^{-2}$)} &\colhead{(erg s$^{-1}$)}
}
\startdata
NGC 1569 &$3.84\times 10^{-11}$ & $1.66\times 10^{40}$ & $2.71\times 10^{-11}$ & $1.17\times 10^{40}$\\
NGC 5253 & $1.76\times 10^{-11}$& $3.37\times 10^{40}$  & $1.67\times 10^{-11}$   & $3.20\times 10^{40}$\\
NGC 4214 &$1.15 \times 10^{-11}$ & $1.19\times 10^{40}$   & $1.00 \times 10^{-11}$    & $1.04\times 10^{40}$\\
He 2-10 &$8.05\times 10^{-12}$ & $7.80\times 10^{40}$ & $7.51\times 10^{-12}$   & $7.27\times 10^{40}$\\
NGC 3077 & $5.28\times 10^{-12}$ &  $9.22\times 10^{39}$ & $3.79\times 10^{-12}$   & $6.62\times 10^{39}$\\
NGC 5236  &$8.06\times 10^{-11}$ & $1.93\times 10^{41}$ & $1.76\times 10^{-11}$   & $4.20\times 10^{40}$\\
  \cline{1-5}\\
Holmberg 2 &$1.77\times 10^{-12}$ & $2.43\times 10^{39}$ & $8.18\times 10^{-13}$ & $1.13\times 10^{39}$\\
NGC 4449 &$3.45\times 10^{-11}$  &$7.32\times 10^{40}$   & $1.79\times 10^{-11}$ & $3.81\times 10^{40}$\\
\enddata
\tablenotetext{a}{Total \has flux with dust extinction correction from each \has image. 
For NGC 5236, we use the central section, A1, from H11, not an entire image.}
\tablenotetext{b}{Total \has luminosity from the \has flux in the first column.}
\tablenotetext{c}{Total \has flux with dust extinction correction from each diagnostic diagram. }
\tablenotetext{d}{Total \has luminosity from the \has flux in the third column.}
\end{deluxetable}

\begin{turnpage}
\begin{deluxetable}{lrrrrrrrr}
\tabletypesize{\tiny}
\tablecolumns{9}
\tablewidth{0pc}
\tablecaption{Measured and Derived Quantities from Diagnostic Diagrams}
\tablehead{
\colhead{Galaxy} 
& \colhead{$L_{H_\alpha,sh}$\tablenotemark{a}} 
& \colhead{SFR\tablenotemark{b}} 
& \colhead{SFR$_{diag}$\tablenotemark{c}} 
& \colhead{$\Sigma_{SFR, HL}$\tablenotemark{d}} 
& \colhead{$\Sigma_{diag}$\tablenotemark{e}} 
& \colhead{ $L_{H_\alpha,sh}/L_{H_\alpha,tot}$\tablenotemark{f}} 
& \colhead{ $L_{H_\alpha,sh}/L_{H_\alpha,tot\_diag}$\tablenotemark{g}} 
& \colhead{ $A_{sh}/A_{tot\_diag}$\tablenotemark{h}} \\
\colhead{}  
& \colhead{(erg s$^{-1}$)}  
& \colhead{($M_{\odot}$ yr$^{-1}$)} 
& \colhead{($M_{\odot}$ yr$^{-1}$)}
& \colhead{($M_{\odot}$ yr$^{-1}$ kpc$^{-2}$)} 
& \colhead{($M_{\odot}$ yr$^{-1}$ kpc$^{-2}$)}
&\colhead{} &\colhead{} &\colhead{} 
}
\startdata
NGC 1569 
& $3.29\times 10^{39}$ & 0.13 &0.067 & 9.6 & 1.5 & 0.20 &  0.26 &  0.28\\
NGC 5253     
& $4.69\times 10^{39}$& 0.26  & 0.11 & 10 & 0.35 & 0.14 &  0.15  &    0.38\\
NGC 4214  
& $1.86\times 10^{39}$ & 0.09 &   0.067 & 2.3 & 0.26 & 0.16  & 0.18   &  0.46\\
He 2-10   
& $1.74 \times 10^{39}$&  0.62  &    0.56 & 33 & 1.4 & 0.022 &  0.024  &   0.28\\
NGC 3077   
& $2.93\times 10^{38}$& 0.073  &  0.052 & 2.6 & 0.42 & 0.034 &  0.048  &    0.21\\
NGC 5236     
& $1.90\times 10^{40}$ & 1.52 & 0.18 & $100>$ & 2.3 & 0.097 &  0.43  &  0.54 \\
  \cline{1-9}\\
Holmberg 2 
& $7.20\times 10^{38}$& 0.019 & 0.0032 & 1.4 & 0.20 & 0.29 &  0.64 &  0.54\\
NGC 4449 
& $6.86\times 10^{39}$ & 0.58 & 0.25 & 4.6 & 0.67 & 0.094 & 0.18 & 0.18\\
\enddata
\tablenotetext{a}{Total \has luminosity with dust extinction correction, associated with shock--ionized gas.}
\tablenotetext{b}{SFR derived from the total \has luminosity in the second column of Table 3.}
\tablenotetext{c}{SFR derived from the \has luminosity in the diagnostic diagram. We 
subtract the shock--ionized component too for the \has luminosity.}
\tablenotetext{d}{SFR density defined as $\Sigma_{SFR, HL} \equiv$ SFR/$A_{SFR, HL}$, where  
the half-light area, $A_{SFR, HL}$, is the pixel area above the half-light surface brightness. 
For NGC 5236, we use the A1 section from H11. Because the section covers the bright central 
region, the half-light flux is higher than the level from the entire image. Hence, the $A_{SFR, HL}$ for NGC 5236 is underestimated. 
The estimated SFR density is several hundreds.}
\tablenotetext{e}{SFR density defined as $\Sigma_{diag} \equiv$ SFR$_{diag}$/$A_{tot\_diag}$,  
where the $A_{tot\_diag}$ is the pixel area of all data points in the diagnostic diagram. }
\tablenotetext{f}{Luminosity ratio to the total \has luminosity }
\tablenotetext{g}{Luminosity ratio to the total \has luminosity in the diagnostic diagram.}
\tablenotetext{h}{Area ratio of the shock--ionized component to the area of all the pixels on the diagnostics.}
\end{deluxetable}
\end{turnpage}

\begin{deluxetable}{crrrr}
\tabletypesize{\footnotesize}
\tablecolumns{5}
\tablewidth{0pc}
\tablecaption{Correlation Test$^\dagger$}
\tablehead{
\colhead{Relation\tablenotemark{a}}& \colhead{Data\tablenotemark{b}}& \colhead{Pearson\tablenotemark{$\dagger$}} 
&\colhead{Spearman\tablenotemark{$\dagger$}} 
& \colhead{Kendall\tablenotemark{$\dagger$}} 
}
\startdata
$L_{sh}$ vs. SFR   & $-$ NGC 1569 & 0.994 & 1.000 (0.000) & 1.000 (0.042)  \\
  & $+$ NGC 1569 &  0.963 & 1.000 (0.000) & 1.000 (0.014) \\
\cline{1-5}\\
$L_{sh}$ vs. $\Sigma_{HL}$ & $-$ NGC 1569 &  0.844 & 0.800 (0.200) & 0.667 (0.174)  \\
  & $+$ NGC 1569  & 0.829 & 0.700 (0.188) & 0.600 (0.142) \\
\cline{1-5}\\
$L_{sh}/L_{tot}$ vs. $M_H$  & $-$ NGC 1569 & 0.917 & 1.000 (0.000) & 1.000 (0.042) \\
  & $+$ NGC 1569 &  0.898 & 0.900 (0.037) & 0.800 (0.050)\\
\enddata
\tablenotetext{a}{The relations shown in Figure~\ref{fig:finalthirteen} and equations~3--6.} 
\tablenotetext{b}{The sub-solar galaxies, NGC 5253, NGC 4214, Holmberg II, and NGC 4449. 
Because no exact correction is available to NGC 1569 due to shallower observation, 
we take the $L_{sh}$ derived from the apparent $5\sigma$ threshold cut  
and perform the correlation tests on both cases of ``with $(+)$'' and ``without $(-)$'' NGC 1569.} 
\tablenotetext{$\dagger$}{We perform the three widely used correlation tests, Pearson, Spearman, and Kendall 
(e.g. Teukolsky et al 1992). The two-sided significances of their deviation from zero for Spearman and Kendall tests are given 
in round brackets, which are values between 0 and 1. A smaller value of significance indicates a more significant correlation. 
}
\end{deluxetable}

\end{document}